\documentclass[twocolumn]{aastex63} 
\usepackage{amsmath}
\usepackage{mathptmx}
\usepackage{txfonts}
\usepackage{graphicx}
\usepackage{placeins}
\usepackage{multirow}
\usepackage{array}
\usepackage{natbib}
\usepackage{tabularx}

\usepackage[T1]{fontenc}
\usepackage{ae,aecompl}
\usepackage{lineno}
\usepackage{listings}

\usepackage{layout}

\usepackage{graphicx}
\graphicspath{ {./images/} }

\shortauthors{Hansen et al.}

\begin{document}
\title[Single Transit Detection in Kepler]{Single Transit Detection In \emph{Kepler} With Machine Learning And Onboard Spacecraft Diagnostics}

\author[0009-0002-7768-620X]{Matthew T. Hansen}
\altaffiliation{matthew.hansen@ufl.edu}
\affiliation{Department of Astronomy, University of Florida, Bryant Space Science Center, Stadium Road, Gainesville, FL 32611, USA }

\author[0000-0001-7730-2240]{Jason A. Dittmann}
\affiliation{Department of Astronomy, University of Florida, Bryant Space Science Center, Stadium Road, Gainesville, FL 32611, USA }


\begin{abstract}
\noindent Exoplanet discovery at long orbital periods requires reliably detecting individual transits without additional information about the system. Techniques like phase-folding of light curves and periodogram analysis of radial velocity data are more sensitive to planets with shorter orbital periods, leaving a dearth of planet discoveries at long periods. We present a novel technique using an ensemble of Convolutional Neural Networks incorporating the onboard spacecraft diagnostics of \emph{Kepler} to classify transits within a light curve. We create a pipeline to recover the location of individual transits, and the period of the orbiting planet, which maintains $>80\%$ transit recovery sensitivity out to an 800-day orbital period.  Our neural network pipeline has the potential to discover additional planets in the \emph{Kepler} dataset, and crucially, within the $\eta$-Earth regime. We report our first candidate from this pipeline, KOI 1271.02. KOI 1271.01 is known to exhibit strong Transit Timing Variations (TTVs), and so we jointly model the TTVs and transits of both transiting planets to constrain the orbital configuration and planetary parameters and conclude with a series of potential parameters for KOI 1271.02, as there is not enough data currently to uniquely constrain the system. We conclude that KOI 1271.02 has a radius of 5.32 $\pm$ 0.20 $R_{\oplus}$ and a mass of $28.94^{0.23}_{-0.47}$ $M_{\oplus}$. Future constraints on the nature of KOI 1271.02 require measuring additional TTVs of KOI 1271.01 or observing a second transit of KOI 1271.02.
\end{abstract}

\section{Introduction}
Since the discovery of the first exoplanets, there has been a rapid increase in the number of exoplanets discovered \citep{Wolszczan_1992, Mayor_1995, Charbonneau_2000}. 
With the discovery of more exoplanets, it became possible to perform demographic studies of exoplanets and dissect the population along other axes (such as stellar metallicity, for example). 
From May 2009 to May 2013 \emph{Kepler Space Telescope}, henceforth \emph{Kepler}, pointed at a single patch of sky and observed roughly 150,000 stars photometrically during its main mission \cite{Borucki_2010}. 
\emph{Kepler} continued to observe the sky after two of its reaction wheels broke as the \emph{K2} mission \cite{K2}.
\emph{Kepler} was a statistical mission aimed at finding the frequency of Earth-like planets around Sun-like stars, $\eta$-Earth.
The Transiting Exoplanet Survey Satellite, (\emph{TESS}), was launched in 2018 and its prime mission lasted until 2020, and now is currently in an extended mission.
TESS is an all-sky survey designed to detect the transits of small exoplanets around the brightest stars \cite{TESS_Ricker}. The primary goal of TESS is to identify transiting planets around bright stars to enable follow-up spectroscopy studies to measure masses and potentially detect their atmospheric composition.

Thousands of confirmed planets and thousands of more planet candidate signals have been found within the \emph{Kepler} field of view \citep{Borucki_2011, Batalha_2013, Robovetter, Morton_2016} as well as within the current \emph{TESS} sample \cite{Guerrero_2021}.
These discoveries have enabled statistical studies of exoplanets, most prominently the occurrence rates of planets as a function of planet size and planet orbital period. 
Of particular interest is the occurrence rate of Earth-like planets around Sun-like stars (i.e. - $\eta$-Earth) 
\citep{Fressin_2013, eta1,eta2,eta3,eta4,eta5,eta6,eta7,eta8,eta9, Hsu_Occurrence,eta11,eta12}.
Although thousands of exoplanets have been discovered within \emph{Kepler}, the sensitivity of \emph{Kepler} requires one to extrapolate out towards the Earth-like parameter space to determine the occurrence rates of small-radii, long orbital period planets, leading to significant variation between the studies on the value of $\eta$-Earth. 

Along with the light curve data of the $\sim$150,000 stars that \emph{Kepler} observed, \emph{Kepler} also produced onboard spacecraft diagnostics throughout its 4-year time span. 
These data include the temperature measurements on sensors throughout the spacecraft, the state of the four reaction wheels, and attitude error pointing in the x-, y-, and z-directions. 
The ancillary engineering data was originally used in detrending and removing systematic errors within the \emph{Kepler} light curves \citep{Twicken_2010a, Twicken_2010b}.
However, using the ancillary engineering data for detrending resulted in significant issues, such as the overfitting of the light curves leading to the removal of stellar variability. 
Cotrending Basis Vectors (CBVs), derived from the collective behavior of all \emph{Kepler} stars, were used instead of the engineering data to detrend the light curves as the CBVs were able to preserve the stellar signal.
Since then, no study has included the ancillary engineering data in their analysis or techniques for exoplanet detection.
The ancillary engineering data may represent one of the last untapped avenues that have not been fully explored when detecting exoplanets within the \emph{Kepler} data set.

Planet discovery techniques like the Box Least Squares (BLS) and periodogram techniques fundamentally rely on phase folding data to stack multiple transits together \citep{BLS, Jenkins_2010}. 
This inherently biases itself towards shorter orbital period planets, where there are more transits, and the signal can become more pronounced. 
Moreover, as the distance between a planet and its host star increases, the geometric probability of the planet transiting decreases.
Therefore, at longer periods where only one or two transits might be visible, these methods are liable to be incomplete.

Further confounding the field of planet discovery, there exist many signals within the star's light curve that can produce a false detection.
False alarms can arise from systematic effects that fail to be taken out of the data, producing threshold-crossing events that upon further inspection fail to be classified as a transit.
False positives can arise from various sources, such as starspots mimicking the behavior of a transiting planet or a background eclipsing binary blending into the light curve appearing as a planet around the observed star.
When applying the BLS method, one way to decrease the number of false positives is to require at least three self-consistent transits \cite{Robovetter}.
However, for an Earth-like planet with an orbital period of 365 days, there are a maximum of 4 transits within the \emph{Kepler} dataset. 
Therefore, using BLS disfavors the discovery of long orbital planets. 
 
By phase-folding the data, one can increase the signal-to-noise ratio SNR of the signal and increase the confidence of a true exoplanet. 
However, for long orbital period planets, phase folding may not be viable as only a single transit of the planet appears within the data.
Therefore, the SNR of one transit is typically the SNR of the signal. 
When searching for long orbital period planets with few transits, one can relax the three self-consistent requirements in favor of a higher SNR \cite{Robovetter, Foreman-Mackey_Long}.
Determining occurrence rates of small radii, and long orbital period planets requires overcoming these two technical challenges in exoplanet detection. 

Of the thousands of planets discovered, a subset of them orbit in mean motion resonances or close enough in orbits that gravitational perturbations can affect their transit times significantly. 
These Transit Timing Variation (TTV) systems can encode information about their masses and eccentricities in this TTV signal \citep{Agol_2005, Holman_2005}. 
Planets that exhibit TTVs do not have linearly separated transit timings but exhibit a periodic behavior.
By observing the precise locations of a planet's transits, one can calculate the TTVs for each transit and the overall periodic signal.
There have been numerous studies to estimate the TTVs of candidates within \emph{Kepler} \citep{Ford2012, Mazeh_2013, Holczer_2016, Kane_2019}.
The studies cataloged the transit timings of thousands of \emph{Kepler} Objects of Interest (KOIs) and showed that on the order of hundreds of KOIs experienced TTVs. 
\cite{Kane_2019} showed that for planets with a period between 3-50 days and a radius between 1.3-6 $R_{\oplus}$, the frequency of strong TTVs increased.

One can utilize the appearance of TTVs within a system with only one transiting planet to infer the existence of a second, non-transiting planet within the system \cite{Agol_2005, Holman_2005}. 
The transit probability of an exoplanet is small compared to the total amount of ways the planet can be aligned. Therefore, it is expected that the vast majority of exoplanets are in non-transiting configurations.
TTVs pose a unique ability to infer the existence of such planets without the need for a transit.
Studies have detected such non-transiting planets within \emph{Kepler} through an in-depth analysis of transiting planets' TTVs \citep{Ballard_TTV, Nesvorny_2012, Nesvorny_2013}. 
\cite{Ballard_TTV} discovered a non-transiting planet through detailed observations and analysis of the TTVs of a 2.2 $R_{\oplus}$ planet.
However, one needs to observe a complete period of the TTV signal of the transiting planet to perform such an analysis.
One can also use the existence of a TTV of a known planet to help validate another planetary candidate within the system. 
Since the existence of a TTV is already indicative of another planet, the probability of a second planet being a false positive decreases.
TTVs also provide additional dynamical constraints when modeling the multi-planetary system, since not every orbital configuration of planets produces the same TTVs.

Machine learning poses a unique skill set that is primed to overcome the challenges of detecting single transit events within \emph{Kepler}.
One can train machine learning algorithms on validated low SNR transiting planets to learn features associated with such planets on a single transit basis.
Machine learning has already been applied to \emph{Kepler} in a variety of ways, such as auto-vetting candidate detections, including Robovetter \citep{Robovetter}, Autovetter \citep{Autovetter}, and Astronet \citep{Shallue_18}.
These auto vetters showed promising results of applying machine learning to validate candidates.

Using convolutional neural networks (CNNs), previous studies were able to train networks on transit-like shapes within phase-folded light curves to accurately predict whether a signal was a false positive or a true exoplanet candidate \cite{Shallue_18}.
\cite{Teachey_21} used CNNs on simulated single transits with injected moon signals within \emph{Kepler} and were able to achieve high accuracy in the classification of simulated exomoon signals. 
Although no exomoons were found within the \emph{Kepler} dataset, they showed that their networks were able to learn the shape of transits within \emph{Kepler} on a single-transit basis. 
Additionally, \cite{Teachey_21} used an ensemble of 50 uniquely built CNNs to classify exomoon signals and found that the accuracy of an ensemble is greater than the accuracy of any one individual network. 

In this work, we create a single transit detection pipeline using an ensemble of CNNs trained on small radii, and long orbital period planets. 
We incorporate the ancillary engineering data as input into our ensemble as well as the photometric data. 
We report the discovery of an additional planet within the KOI 1271 system and perform a series of tests to constrain the planetary and orbital parameters.

In Section \ref{sec:NN} we describe our methodology for building the data set as well as the CNN architectures. 
In Section \ref{sec:pipeline} we describe our end-to-end pipeline from going from a star with no prior knowledge to the locations of the planetary transits and its period. 
In Section \ref{sec:Candidate} we demonstrate our pipeline on a known system, report the discovery of a single transit candidate, and give a suite of solutions for its orbital and planetary parameters.
We conclude in Section \ref{sec:Conclusion}.

\section{Neural Network Implementation} \label{sec:NN}
\subsection{Network Training Set}
We assume, a priori, that not every discovered planet will be beneficial in training a machine learning framework to discover small radii, and longer orbital period planets to push the sensitivity threshold of \emph{Kepler}. We therefore begin the data set processing by making a series of cuts to the planetary candidate sample. The two cuts we make are: 1) an orbital period greater than 12.5 days, and 2) a transit depth less than 350 parts per million (ppm).
The input into our network must be equal to 500 points, explained in \S\ref{sec:NN_design}. 
As we only use long-cadence data in our implementation, 500 long-cadence points are roughly 10.5 days. We include an extra 2 days in our down-selection process to have a little leeway and ensure no overflow of one segment into the next.
Since no transits appear in the training set more than once, we ensure that a single transit is not over-represented in training and the network does not overfit.
A depth of 350 ppm roughly corresponds to a 2$R_{\oplus}$ planet around a 1$R_{\odot}$ star. Our two cuts therefore limit our sample of candidate / confirmed planets to long orbital planets of roughly Earth size. Our final training, validation, and testing set contains 621 planetary signals with 546 unique systems, down from an original 4033 planets with 3068 unique systems \cite{koidr25}.

The goal of our single transit detection network is to state whether a 500 datapoint light curve segment contains a transit or not, in other words, our problem is a binary classification problem. 
Our dataset must resemble a binary classification problem, and have two labels: 1) Transit and 2) No Transit. 
We produce a dataset containing each individual transit available from the sample candidates, along with a dataset containing 10-day segments of stars' light curves with no transit within them. 
To grab the dataset containing known transits, we use the Mikulski Archive for Space Telescopes (MAST) database to obtain the ephemeris of each known planet. 
We place a 500-cadence window centered on the transit and check if the initial and final time stamp of the window is within 11.5 days of each other, about a 10\% leeway, to ensure there are no large data gaps within the data segment. 
The flux corresponding to each timestamp is collected and stored in an array, along with the time stamps, the \emph{Kepler} Input Catalog (KIC) number of the star, and the name of the planet. 
Also for each window segment, we collect a series of on-board spacecraft diagnostics associated with each time stamp. 
For a full list of attributes collected see \S\ref{sec:NN_design}. 
However, since the time stamps listed in the light curve are not a one-to-one correspondence with the time stamps in the engineering files, due to a shorter cadence for the onboard spacecraft diagnostics, we average the engineering files over the 30-minute cadence associated with the flux values. 
The final size of the array for a single 10-day window segment is (20502).
The above procedure is repeated for every transit for each planet in our down-selected sample. 
We do not check if there are other transits within the 10-day window for the in-transit set, since if there are more transits within the window segment the network will have more features to be trained on in classifying a transit.

The no-transit sample poses a unique set of problems. 
We must ensure that there are no transits within the 500 data points for the no-transit sample. 
This requires that a given star has consecutive 500 data points where there is no transit.
However, there are stars that do not possess consecutive 500 data points without a transit.
For stars with multiple planets, we flag any time stamps pertaining to any one of the transiting planets and ensure that the 10-day no-transit segment contains no flagged time stamps.
We then start on the zeroth index of a given star's light curve and check if any of the time stamps between the initial one and the 500 one are not flagged as belonging to a transit and if the total time is within 11.5 days. These two conditions must be true for data to be grabbed and put into the no-transit set for training, validation, and testing. 
Therefore, it is easy to notice that there are stars within our down-selected sample that produce no data for the no-transit sample. Since we require a 10-day window of data containing no transits, but there are some planets in our sample with a 12.5-day period, those planets will not produce any out-of-transit segments. There are also some long-period planets that if they were single planetary systems would produce some no transit segments, but since there are other planets within the system with short periods no data is produced for them. 
The last check we make is to make sure that we have not collected a no-transit sample for that star before. Since some stars have multiple planets that pass our selection process, we do not want to double the no-transit data for that star.
We are still able to produce a complete dataset with a 50/50 split between transit / no transit since a single planetary system with a long orbital period produces a plethora of no-transit data segments.
We note here that some of the no-transit data segments may still be mislabeled, since they may contain transits that have not yet been discovered. 
There is no way to prevent this since these planets are undiscovered and the goal of the project.
Since we are constructing tens of thousands of no-transit segments it is unlikely that there are undiscovered transits in a large portion of the training set to skew our training.
Therefore, we can conclude that this is statistically insignificant in the long run and move forward.

The final sizes of the transit dataset and no transit dataset are (26066, 20502) and (35627, 20502), respectively. 
We implement a randomized 80/10/10 training, validation, and testing split. 
To make sure that every star only contributed to one of the training, validation, and testing sets we randomized the split based on the stellar ID. 
By doing this, we ensure that the network is not biased by any particularities associated with any one specific star. 
There is a difference of 9,000 data segments between the transit dataset and the no-transit dataset; however, for classification problems, it is best to have a 50/50 split so the network does not prefer one classification over the other.
To make a 50/50 split within each training, validation, and testing split we randomly down-select the no transit portion to be equal to its transit counterpart's size. This will allow the network to train appropriately and give an accurate depiction of the testing results. 
The amount of data segments in the training, validation, and testing are 41818, 5046, and 5268, respectively, all with a 50/50 split in their labels between transit and no transit. 

\subsubsection{Data Standardization}\label{sec:NN_data}
One of the most important parts of machine learning, and neural networks particularly, is the way the data is treated before being inputted into the network. The flux values obtained by \emph{Kepler} need to be normalized in some way so that each star can be treated equally and the network can be set up for success in picking out transit-like shapes in the light curve. 
We start the process of editing the flux values by using \texttt{LightKurve}'s normalization function which divides the whole light curve by the median on a quarter-by-quarter basis \cite{Lightkurve}. 
However, each star has its own unique activity profile due to a variety of causes such as star spots. 
These unique profiles can actually mimic transits and influence the training of the networks to pick up on non-transit shapes, therefore we must also detrend each light curve independently. 
We utilize a best-fit spline, which uses b2 statistics, to detrend the light curves \cite{Vanderburg_b2}. Importantly, we do not mask the transits when we detrend the light curves nor do we give any information to the spline about where the transits are located. 
The process of not masking transits from the spline is vital since it treats each star how we would treat a star when searching for an unknown planet, not knowing where the transits are located.
The next step after detrending the light curves is to normalize the data segments themselves which are the inputs into the network. 
We use the same normalization for the flux as \cite{Teachey_21} using the given equation

\begin{align}
    {\bf F}_{\rm{norm}} &= \frac{{\bf F} - \rm{min}({\bf F})}{{\bf\tilde{F}} - \rm{min}(\bf{F})} - 1,
\end{align}
where $\bf F$ is the array of fluxes, $\bf \tilde{F}$ is the median value of the flux input, and min($\bf F$) is the minimum of the flux array. We subtract 1 to center all the fluxes about 0. Early testing in trying different forms of normalization showed that the above was the best one. In fact, we found some forms of normalization inhibited the network from learning anything other than random guessing.

For the engineering attributes, we perform a standardization on each individual attribute. We find the median value for all data points in each attribute, as well as the median absolute deviation (MAD). We then take the difference between each individual attribute's data point and the median and divide the difference by the MAD, hence turning each point into the distance from the median in units of MAD.

\subsection{Neural Network Design} \label{sec:NN_design}
We choose to work with convolutional neural networks (CNNs) instead of regular neural networks (NNs) due to the nature of the problem, as we are trying to detect the shape of a transit, which CNNs are optimized to do. We use \cite{Shallue_18} and \cite{Teachey_21} as a basis for our CNN structures, where we use a series of convolutional layers, and pooling followed by a fully connected neural network (FCNN) to classify transit or no transit. 

Previous works, \cite{Teachey_21}, have shown that an ensemble approach can be used to increase the overall accuracy of a framework compared to a single network. Although \cite{Teachey_21} used an ensemble approach to try to detect signals from exomoons, we find similar results that using an ensemble greatly increases the accuracy compared to a single network. The ensemble approach is beneficial if each network in the ensemble either has a unique structure or a unique training set so each network can hopefully learn different traits relating to a correct classification. We choose to build an ensemble of CNNs trained and tested on the same datasets but differ in their unique structure. We explain how we find the unique structures for each network below.

\begin{figure}
    \centering
    \includegraphics[height= 10 cm]{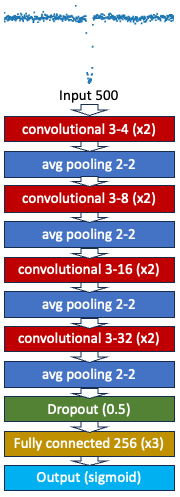}
    \caption{One of the 25 architectures within the ensemble using only the flux values of the host star. The convolutional layers are described with \emph{kernel size - \# of filters, (\# of layers)}, pooling layers are described with \emph{pool size - stride length}, and the fully connected layers with \emph{\# of neurons (\# of layers)}. Shown above is an example input of a detrended light curve, containing 500 long-cadence data points.}
    \label{fig:NN_arch}
\end{figure}

We explore a range of possible hyperparameter combinations to determine the best structures to use in our approach. With the help of UF's Hipergator resources, we were able to do a manual grid search over each possible combination of the hyperparameters we chose to tune. The hyperparameters we performed a grid search over were along with their range of values: 1) Pool Type (max, average); 2) Pool Size (2, 3, 4, 5); 3) Stride Length (1, 2); 4) Convolutional Layers (1, 2, 3, 4, 5); 5) Dropout Rates (0.0, 0.25, 0.5); 6) Dense Layers (1, 2, 3, 4); 7) Neurons (128, 256, 512). 
A sample architecture is shown in Figure \ref{fig:NN_arch}.
We trained and tested each possible combination of these hyperparameters, using just the flux data, creating a total of 2880 unique networks. For training purposes, the batch size was held at a constant 128, as early testing showed that was the best option for speed and results of training networks. 
A kernel size of 3 was used for each convolutional layer. We adopted \cite{Shallue_18} notation in that the convolutional layers came in pairs, meaning when we state 1 convolutional layer, it is actually 2 layers in sequence followed by a pooling. 
We use a sigmoid function for the final output, which makes the output a proxy for how confident the network is in its classification. A value closer to 1 represents strong confidence in the prediction of a transit, and a value closer to 0 represents strong confidence the segment contains no transits. Figure \ref{fig:NN_arch} shows one of the architectures out of the 2880 that were tested.

\subsubsection{Confidence Metric}
We must combine each network's confidence in some way to produce a single value for the overall ensemble's confidence.
There are many schools of thought of what is the best way to combine the results in an ensemble method for machine learning, some even use another NN to find the best weights for each individual network to combine them into one final result \citep{ens_ml, ens_alg_ml}. 
Here we take the average of all the individual network's output to create a single value representing the overall confidence of the ensemble. 
A value close to 1 represents a strong confident agreement across the ensemble that there is a transit within the segment. 
A value close to 0 represents a strong confident agreement across the ensemble that there is no transit within the segment.
A value close to 0.5 can represent a few scenarios: either the individual networks have no confidence in their prediction making the ensemble have no confidence, or half of the networks are strongly confident that there is a transit, and the other half that there is no transit. Either scenario leads to the same results: the ensemble as a whole is not confident in its classification.

We state an arbitrary threshold of 0.5, in which the confidence metric must be greater than to result in a transit classification. 
Changing this threshold can tune the precision and accuracy of the network, and early testing found that a threshold of 0.5 gave the optimal chance of recovering transits while also being selective in its classifications of transits.

\subsection{Neural Network Results} \label{subsec:NN_results}
\subsubsection{Flux Only}
We trained and tested each network structure on the same training, validation, and testing dataset. We trained each structure to 100 epochs, but with an early stopping condition that if the validation accuracy did not improve over the last 5 epochs the training was stopped. After each network was trained, we tested the final network on the testing set and recorded its accuracy. 
We observed that some architectures performed significantly worse than the average and a few architectures even struggled to perform better than random chance.

\begin{figure}
    \centering
    \includegraphics[width = 8cm]{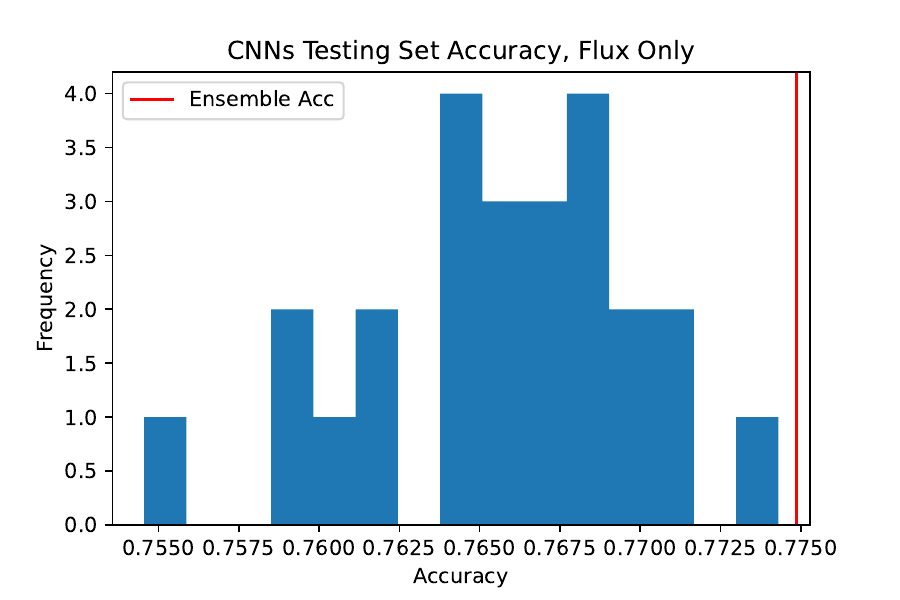}
    \caption{Histogram of the individual accuracies of the 25 networks in the flux-only ensemble. Also shown is the ensemble accuracy. The ensemble accuracy is greater than any one individual accuracy within the ensemble.}
    \label{fig:flux_accuracies}
\end{figure}

We saved the top 25 performing networks to create our ensemble, all with an accuracy of over 75\%.
Although \cite{Teachey_21} used 50 networks in their ensemble, early testing showed little improvement when increasing from 25 to 50 networks in the ensemble but a significant increase in total time to compute classifications.
A histogram of their accuracies can be seen in Figure \ref{fig:flux_accuracies}.
Figure \ref{fig:flux_accuracies} shows that the ensemble method is able to significantly improve the accuracy of detecting transits greater than any one individual structure. 
We note here that this flux-only structure serves as a baseline to test the improvement that may be gained from including onboard spacecraft diagnostics.

\subsubsection{Including On-Board Spacecraft Diagnostics}
The exploration of different structures of CNNs only used the obtained flux values in the data sets and ignored the collected onboard spacecraft diagnostic data. This was intentional since there is a significant speed increase in the time it takes to train and test the networks when compared to including the onboard spacecraft diagnostics.

If fluctuations in spacecraft health or temperature can imprint on measured fluxes in the data, then including these values in our classifier may help identify transits previously lost in noise.
The ancillary engineering files are bundled in 8 comma separated value files, representing 6 different engineering attribute categories: 
1) Attitude Errors; 
2) Board Temperatures; 
3) Mount Temperatures; 
4) Optics Temperatures; 
5) Reaction Wheel Speeds; and 
6) Telescope Temperatures. 
Within each of these broad attribute categories, there are more specific attributes (e.g. attitude error of each axis of rotation) totaling 39 specific on-board spacecraft diagnostics. 

\begin{figure}
    \centering
    \includegraphics[height= 10 cm]{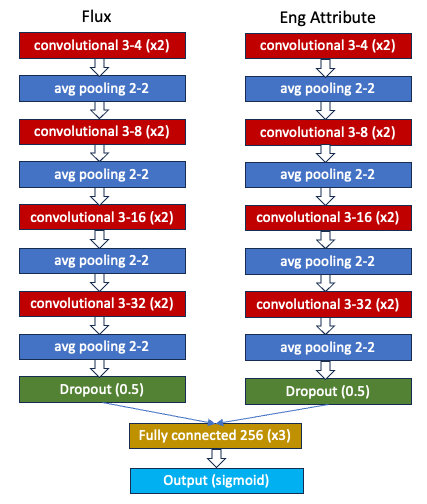}
    \caption{Example of one of the 25 architectures in the specific engineering attribute ensembles used to determine the best-performing attributes. 
    The columns follow the same convention as Figure \ref{fig:NN_arch}. Each column's input size is 500. After the dropout layer, the outputs for the two columns are concatenated and used as input into the fully connected layer. }
    \label{fig:2_col_CNN}
\end{figure}

One can assume that not every diagnostic is as important as the rest, and will help contribute something impactful to a neural network, but we do a simple check to make sure. 
We create, train, validate, and test 39 separate ensembles, each with 25 models, with 2 disjoint columns one taking the flux as its input and another taking a specific diagnostic as its input. 
Figure \ref{fig:2_col_CNN} shows an example CNN architecture for this test.
In addition to the 39 ensembles, we create 14 more ensembles that as the engineering input use the difference between temperature attributes within the same group.
We perform this step since one might assume that maybe it is not the temperature itself that affects the flux, but the temperature differential between different points on the spacecraft (i.e. - a temperature gradient). 
For example, the temperature difference between two sensors on the Schmidt Corrector might impact the observed flux.
For that reason, we added the 14 additional trained ensembles to test what the best attributes to be added are.
Due to limited computational resources and time, we cannot do what we did when deciding the structure of the top neural network models and perform a manual grid search over every possible combination of the diagnostics. 
We use the 53 unique ensembles as a proxy for which diagnostic adds the most additional information to improve the accuracy of the ensemble over using only the flux values.

The baseline accuracy that we use for comparison is that of the ensemble trained using only the flux values, which was 77.5\% seen in Figure \ref{fig:flux_accuracies}. 
Out of the 53 unique ensembles, 40 of them had overall accuracies higher than the flux baseline. 
The only attributes that performed worse belonged to the difference in temperatures of the sensors within the Schmidt Corrector and also the Primary Mirror. 
Out of the 53 ensembles, 11 of them achieved an ensemble accuracy greater than 79\%, and 2 of them achieved an accuracy greater than 80\%.

\begin{figure*}
    \centering
    \includegraphics[height = 8 cm]{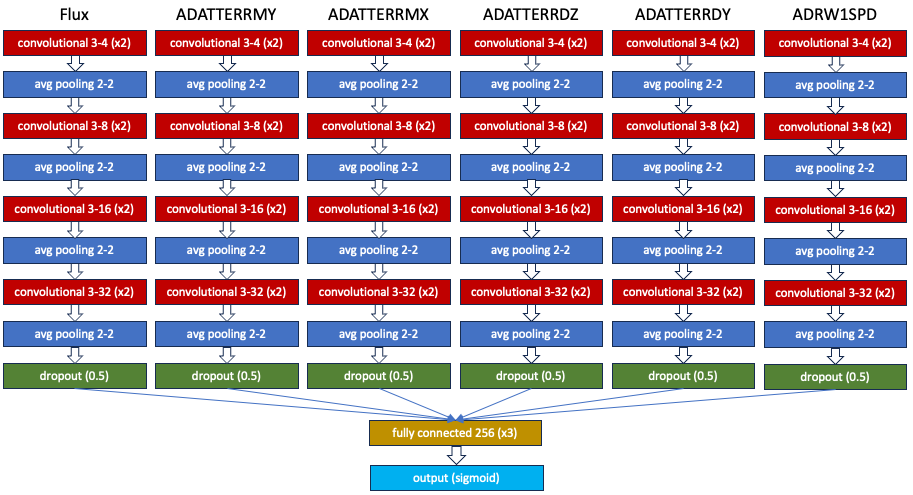}
    \caption{One of the 25 architectures in our final ensemble including the engineering attribute columns. The columns follow the same convention as Figure \ref{fig:NN_arch}. The name of the attribute for each column is listed at the top. Each column's input size is 500. After the dropout layer, the output for each column is concatenated and used as input into the fully connected layer. }
    \label{fig:full_NN}
\end{figure*}

We now have 54 total ensembles, one utilizing only the flux values and 53 utilizing the flux coupled with one individual engineering attribute. 
However, we want one final ensemble that includes the onboard spacecraft diagnostics that increase the accuracy but do not sacrifice too much computational time.
We make two additional ensembles: one ensemble with 25 networks using the top 5 engineering attributes in tandem, and one ensemble with 25 networks using the top 11 engineering attributes in tandem. Figure \ref{fig:full_NN} shows an example CNN architecture within the top 5 ensemble. 
The top 11 attribute ensemble follows the same structure as shown in Figure \ref{fig:full_NN}; however, there are 12 total columns instead of 6.
Each engineering attribute was a separate column within the CNN, and all columns were concatenated and inputted into the FCNN.
The top 5 individually performing engineering attributes are the mean attitude error about the y-axis and x-axis; the standard deviation of the attitude error about the z-axis and y-axis; and the reaction wheel \#1 speed.

\begin{figure}
    \centering
    \includegraphics[width = 8cm]{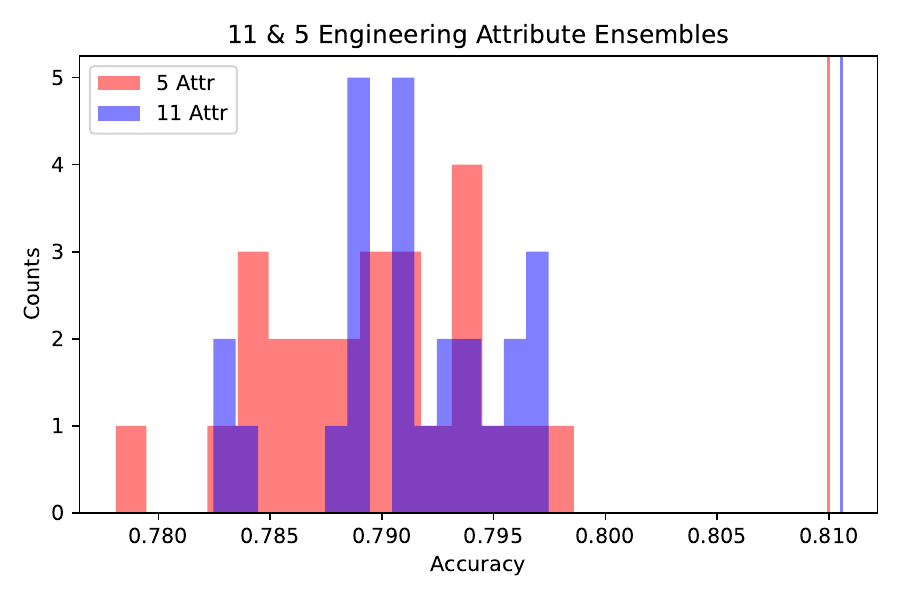}
    \caption{A comparison of two ensembles, one including the top 11 engineering attributes along with the flux (blue) and one including the top 5 engineering attributes along with the flux (red). The histograms are the individual network accuracies within the ensemble, and the vertical lines are the respective ensemble accuracy. The ensemble accuracy for both ensembles is well above any individual accuracy within the ensemble. We also see over a 3\% increase in the ensemble accuracy compared to using only the flux values (see Figure \ref{fig:flux_accuracies}. }
    \label{fig:11_5_attr}
\end{figure}

When we compared the results of the top 5 and top 11 ensembles, there was little difference between the overall ensemble accuracy. 
However, there was a significant time increase when training and testing the top 11 attribute ensemble. 
The ensemble using the top 11 and top 5 attributes had an accuracy of 81.1\% and 81.0\%, respectively. Figure \ref{fig:11_5_attr} shows the comparison between the two ensembles, with the individual CNN performance as well as the overall ensembles' accuracies.
We therefore decided, that the minimal increase in accuracy using 6 more engineering attributes did not warrant the significant time increase and therefore decided to stick with only using the top 5 attributes as our final model, Figure \ref{fig:full_NN} for reference architecture.

\section{Pipeline} \label{sec:pipeline}
We have shown that we are able to achieve high accuracies on labeled data sets of known planets. 
However, we wish to be able to apply our pipeline to every star in \emph{Kepler} without any prior knowledge of whether there is an orbiting planet and discover new planets. 
To feed each star into our ensemble, we use a sliding window function in time that grabs all the associated flux and engineering data corresponding to that window segment. 
This is the exact same method used to grab data as in \S\ref{sec:NN_data}; however, instead of centering the window on transits we start with the first timestamp of a quarter and move forward in time, one timestamp at a time, through all available data. 
Each data point will therefore appear in up to 500 individual light curve segments submitted for classification to the ensemble.
For uniformity of the data and to structure the data as similarly as possible to the training set we impose the same restrictions as in \S\ref{sec:NN_data}, most importantly that the first and last cadence in the window must not be more than 11.5 days apart. 
We found this to be a good leeway from the 10.5-day standard, to where we can still collect enough window segments near data gaps and will also not collect too many window segments that contain large data gaps within them. 
The total number of light segments to classify a single star is then a function of how many 500 consecutive point windows can fit within the light curve while maintaining a total time of less than 11.5 days from the start to the end of the window segment.

\subsection{End-to-End Process of Classifying a Detection}
Once we have collected a series of window segments for a single star, we then need to feed the series into the ensemble, classifying each window as containing a transit or not, recovering locations of the most promising detections, and measuring a period if there is more than one flagged transit candidate. 
We will walk through an example of a star with a known long-orbital period planet outside the training, validation, and testing set. 
 We selected the system KOI 622 as our benchmark example. 
The host star is $2.14^{0.25}_{-0.24}$ $R_{\odot}$, and the known candidate planet has a radius of $16.28^{1.91}_{-1.86}$ $R_{\oplus}$ with a period of 155.04 days \citep{Berger_2023, Holczer_2016}. 
This test case is well outside the initial conditions we used to down-select out training, validation, and testing sets. However, it serves as a good test case since the signal in the light curve is easy to locate and the planet has an average single transit signal-to-noise ratio (SNR) of 41.68. 
The system has a duty cycle of 88\%, meaning out of all the 30-minute intervals between the start and end of \emph{Kepler}, the KOI 622 system has data pertaining to 88\% of the time stamps.

After applying our sliding window segment to collect a series of data on the star, we collected a total of 38,186 lightcurve window segments. 
Each segment contains the stellar ID, the name of the planet, the 500 timestamps associated with the window, the 500 corresponding long-cadence flux values, and all 39 engineering attributes corresponding to each time stamp. 
The final size of the 2D array collected for the 4-year timespan of collected data of the star is (38186, 20502). 
The collected data is then standardized using the same techniques as described in \S\ref{sec:NN_data} and fed into the ensemble of CNNs.

\begin{figure}
    \centering
    \includegraphics[width = 8cm]{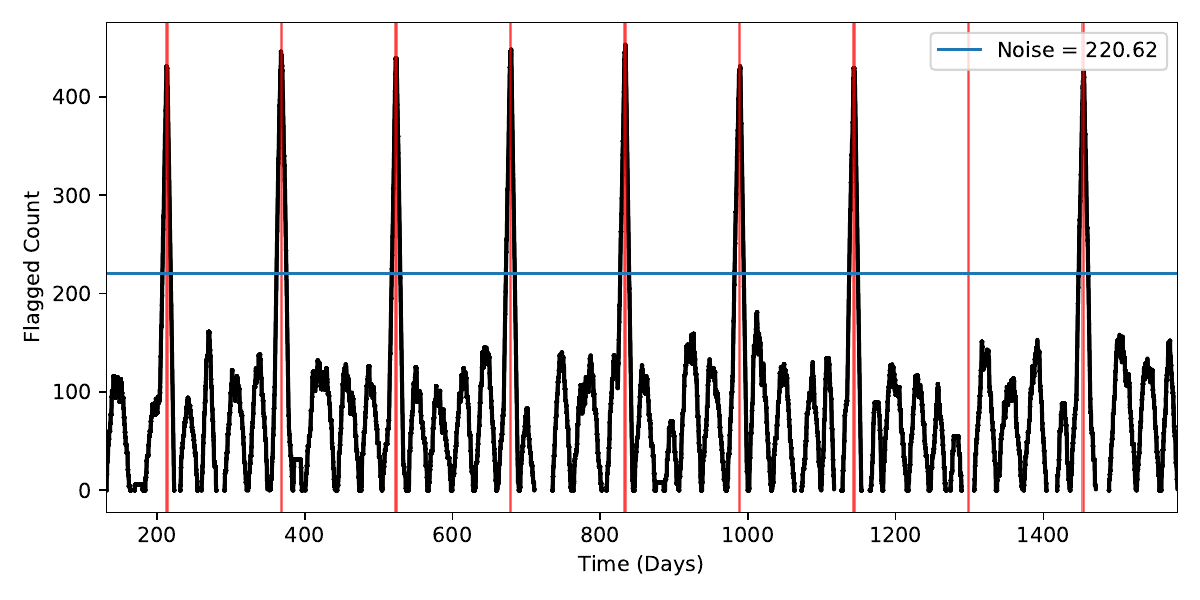}
    \caption{A time series of the total number of times each timestamp was flagged as being within a window segment containing a transit for system KOI 622. KOI 622 has a radius of $2.14^{0.25}_{-0.24}$ $R_{\odot}$. The candidate planet, KOI 622.01, has a radius of $16.28^{1.91}_{-1.86}$ $R_{\oplus}$, and a period of 155.04 days with the first transit appearing in the data at day 213.50. The red highlighted regions are the known locations of the transit events. The second to last transit falls within a data gap of \emph{Kepler}, as noted with no data within the region of the transit. The horizontal blue line is the (arbitrarily) defined noise level, 3 MAD above the MAD of the flagged counts. Our ensemble accurately recovers 100\% of all available transits of the system. We also correctly recovered a period of 155.11 days. This signal would be classified as a successful recovery.}
    \label{fig:test_case}
\end{figure}

We then join the results in the same manner as in \S\ref{subsec:NN_results}, which produces a binary array of length 38186 stating whether the given window segment contains a transit. 
Since a single timestamp can appear in multiple individual window segments, we count the total amount of times a timestamp was within a segment that was flagged as containing a transit. 
Shown in Figure \ref{fig:test_case} is the total amount of times a timestamp was flagged as being within a window segment containing a transit for the whole duration the host star was observed by \emph{Kepler}. 
The highlighted regions are locations of the transits by the candidate planet. 
The second to last transit falls into a data gap, but we keep the region highlighted for clarity of the period. 
Our ensemble does not have 100\% accuracy, therefore we expect every star to have some noise level where the ensemble falsely predicts there is a transit.
We define a false positive noise floor for each star as 3 median absolute deviations (MAD) above the MAD of these values.
The noise level for this star is 220.62 shown in Figure \ref{fig:test_case} as the horizontal blue line. 
Using a peak-finding algorithm, we use the noise level as a minimum threshold for peaks. 
With this defined noise threshold, we recover all 8 of the transits of KOI 622.01 that appear in the \emph{Kepler} dataset.

This method has successfully identified the locations of individual transits within a \emph{Kepler} lightcurve; however, we still need a method to recover the period of the detected planet. 
We are able to run a Lomb-Scargle Periodogram on the flagged counts (Figure \ref{fig:test_case}) and obtain accurate results. 
For our test case, we recovered a period of 155.11 days compared to the reported period of 155.04 days.
The discrepancy is roughly equal to 100 minutes or 3 \emph{Kepler} long cadences.
This method works well when there are few data gaps within the light curve and each time stamp appears in close to 500 individual submitted light curve segments. However, early testing showed that this approach started to fail to recover accurate periods when the duty cycle of stars decreased (i.e. - increased numbers of data gaps).

\begin{figure}
    \centering
    \includegraphics[width=8cm]{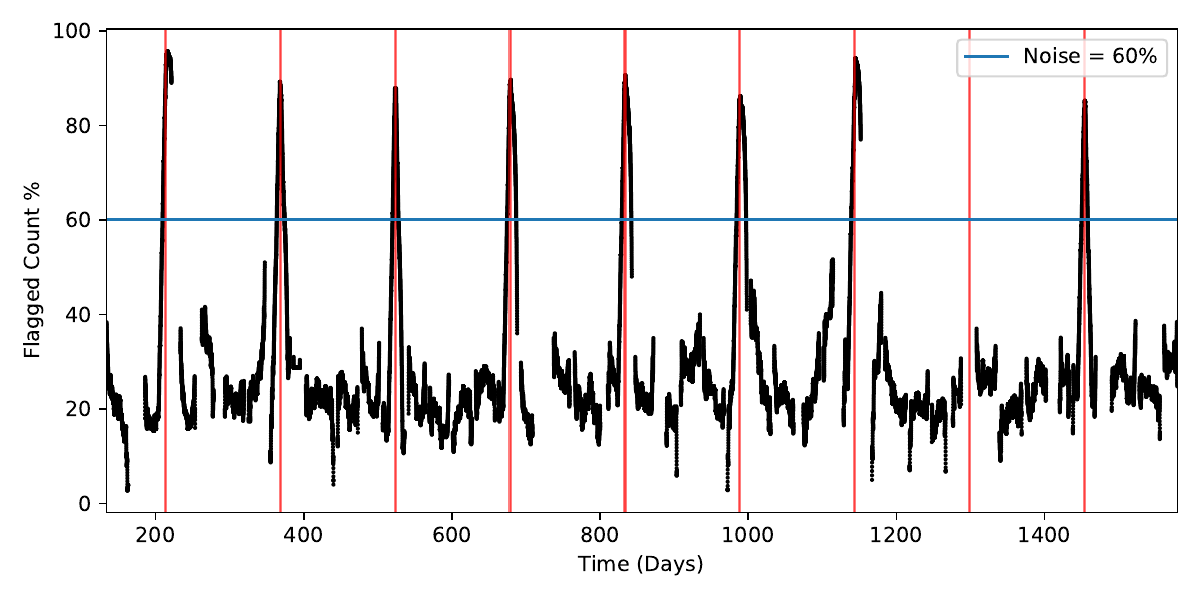}
    \caption{A scatter plot of the percentage each timestamp was flagged as transit over the total amount of times the timestamp appeared within the data set. This is a modified version of Figure \ref{fig:test_case}. We require a timestamp to be in at least 100 light curve segments to be included in the plot. The red highlighted regions are the locations of when KOI 622.01 transits the host star. The second to last transit falls within a data gap. The horizontal blue line is the defined noise level of 60\%. We successfully recover all visible transits above the defined noise level. We correctly recover a period of 155.00 days.}
    \label{fig:percent_flagged}
\end{figure}

When a timestamp is within the 11-day leeway of a data gap, the frequency at which the time stamp appears in individual submitted light curve segments decreases.
As the data gaps grow in size or number, the frequency continues to decrease.
If a transit were to appear near a large data gap, the timestamps for the transit might only appear within the data set a handful of times. 
The ensemble might flag each timestamp pertaining to that transit 100\% of the time the ensemble sees them; however, since they only appeared within the data a handful of times, the \textit{total} flagged count is well below the noise level. 
We therefore adopt an additional metric: the fraction of the amount of times a timestamp was flagged as a transit over the total amount of times the timestamp was fed into the ensemble. 
We add an additional constraint such that a timestamp must appear in at least 100 individually submitted light curve segments.
This is to avoid a scenario where a timestamp appears within the ensemble once or twice but is flagged 100\% of the time.
The final result for the test case KOI 622, when this new metric is applied, is shown in Figure \ref{fig:percent_flagged}.
By adopting this new metric, we can define a new noise threshold of 60\%, meaning that a timestamp needs to have been flagged more than 60\% of the time when the timestamp was fed into the ensemble. 

Once the percent flagged count is created, we can make a pulse train based on the peaks above 60\%. 
We use \texttt{SciPy}'s signal peak finding algorithm and impose that peaks must be above 60\% and at least 10 days apart from one another \cite{scipy}. 
We then create and initialize an array of zeros like the timestamps from the percent flagged count array. 
We find the timestamps associated with each peak and set all timestamps $\pm$ 10 days equal to 1. 
This creates a signal that emulates a pulse train.
We run ASTROPY's Lomb Scargle Periodgram on this modified pulse train. 
For KOI 622, we recover a period of 155.00 days, which is extremely close to the 155.04-day period of the candidate planet.
The discrepancy is 60 minutes or two \emph{Kepler} long cadences, a factor of 1.7 times better than the previous method.

There are a few edge cases where we cannot or do not use the Lomb Scargle Periodogram on the modified pulse train: 1) no peaks above the noise threshold were found; 2) only one peak was found above the noise threshold; 3) two peaks were found. 
In the case of no peaks, we report no detection of an exoplanet and move on.
For the case of a single significant peak found, we run a periodogram on the flagged count array to observe if there is any underlying periodic signal below the noise level that might be attributed to the single transit event. We mark the light curve as a single transit event system for further manual inspection and move on.
When we find only two significant peaks, we take the difference between the two peaks as the estimated period, as we found this to produce better results than using a periodogram dominated by noise.

\subsection{Feature Mapping}\label{subsec:feature_mapping}
The window size used for collecting data is roughly 10 days. When we find peaks within the flagged count plots, see Figure \ref{fig:test_case}, the peaks can be anywhere from $\pm$ 10 days of the actual transit. This is because the transit itself slides from one end of the input window to the next, making the peak anywhere from $\pm$ 10 days from the transit. This is not precise enough for locating a transit and we would require a more precise transit center. Of course, we can manually inspect each peak and find the precise location. However, we anticipate applying our pipeline to the complete \emph{Kepler} data set in future work, which would necessitate an automated process. Moreover, feature mapping will inform us what features in the light curve our ensemble is picking up as a transit. 

To produce a feature mapping plot, we must first decide how to deal with the initial uncertainty of the transit location. 
We do not know where in the potential 20-day peak the transit lies and our ensemble can only take a 10-day input. 
We therefore split the 20-day peak into 3 overlapping sections to be used in the feature mapping process. 
The first section is the first 10 days of the peak; the second section is the middle 10 days; and the third section is the last 10 days.
By creating 3 sections in this way, we increase the likelihood that one of the sections will have the transit within the center of its section.
Early in our testing, we observed that our ensemble prefers the transit to be centered within the window (as our training set was) and the 3 overlapping sections ensure that the transit is near the center in at least 1 window. 

After 3 windows about the peak are selected, we employ the same technique as laid out in \cite{Teachey_21} to perform feature mapping. 
We used a moving filter replacement of 50 points, equivalent to 25 hours, setting the flux and engineering attribute values to 0 within the filter. 
Since there is information contained within the engineering files, setting their values within the filter to 0 was of significant importance for accurate feature mapping.
We slide the moving filter replacement to the right one data point at a time. 
Every time we move the filter replacement, we set the corresponding values equal to 0 and rerun the segment through the ensemble for classification.
We initialized an array of zeros equal to the length of timestamps within the window segment. 
If a timestamp was within the moving filter, if the ensemble still predicted a transit within the window segment, we increased its value by one. If the ensemble no longer predicted a transit, we decreased its value by 1.
Therefore, a timestamp within the transit should have a value close to -50, and a timestamp outside the transit should have a positive value.
We then attribute the lowest value within the array to being the transit center. 
Due to edge effects, we cannot attribute any points within 50 points of the edge of the window to being the transit center, and rely on the 3 overlapping windows to accurately find the transit center if it does occur within one of those points. 

\begin{figure}
    \centering
    \includegraphics[width=8cm]{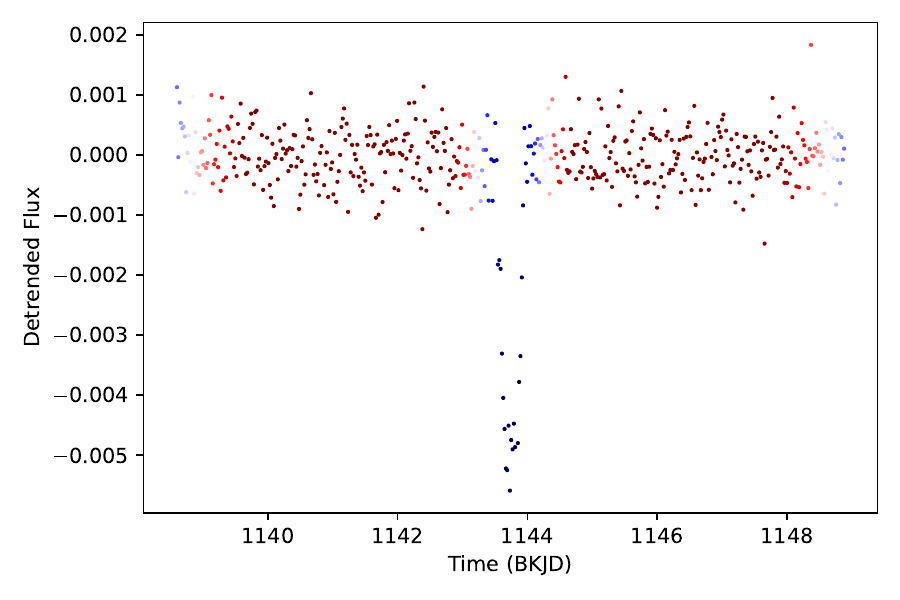}
    \caption{Feature mapping of the ensemble reported transit at 1144.73 BKJD for KOI 622.01. The y-axis is the detrended flux that is the input into the ensemble. The color blue represents the ensemble no longer predicting a transit for the window when those points are mapped to 0. The color red represents a transit classification when that point is set to 0. 
    The newly reported transit center is 1143.61 BKJD, 7.6 times more precise than the ensemble reported value.  
    Our ensemble correctly identifies the features within the light curve that are attributed to the transit. We note the edge effects near the beginning and end of the window.}
    \label{fig:feature_map}
\end{figure}

Figure \ref{fig:feature_map} shows the result of feature mapping the peak at day 1140 BKJD in Figure \ref{fig:test_case}. 
Red points indicate regions where the ensemble still predicted a transit when they were filtered out. 
Blue points represent locations where the ensemble no longer predicted a transit when they were filtered out. 
We also note the edge effects on data within 50 points from either edge. 
One of the transits of KOI-622.01 lies at 1143.74 BKJD.
The ensemble predicted a transit at 1144.73 BKJD.
The result from feature mapping produces a more confident, precise location of the transit center of day 1143.61 BKJD. 
The feature mapping transit center is 7.6 times more precise than the ensemble-reported center.

Although when performing feature mapping we produce a much more precise location of the transit, we choose not to use it when trying to find the period. In testing, we found no improvement in period recovery when using feature mapping compared to using the already found peaks, yet there was a significant time increase to perform the feature mapping. This is a limitation due to the window function of the Lomb Scargle Periodogram and we refer readers to \cite{LombScargle} for further reading on the inner workings of the periodogram. 
Even though feature mapping is not used within the finalized pipeline, we have shown that our ensembles are able to accurately learn the shapes of a single transit and detect them in \emph{Kepler} light curves.

\section{New candidate in the KOI 1271 system} \label{sec:Candidate}
\begin{figure}
    \centering
    \includegraphics[width=8.5cm]{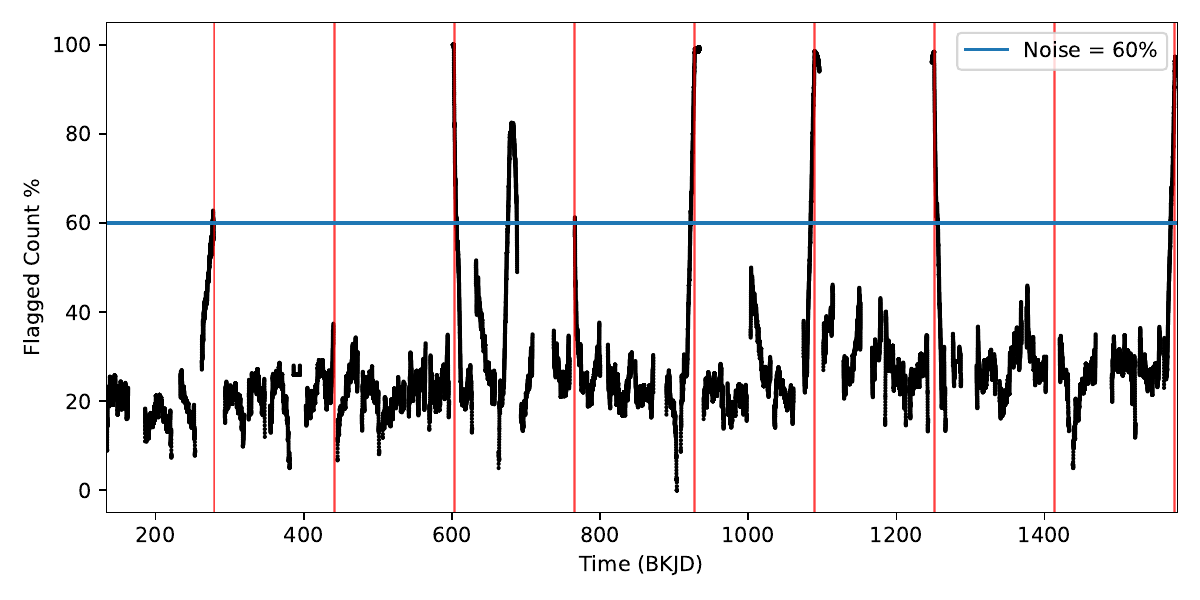}
    \caption{Scatter plot of the percentage each timestamp was flagged as being in transit of KOI 1271. The star is $1.48^{0.09}_{-0.07}$ $R_{\odot}$ with one known planet, KOI 1271.01, with a radius of $11.23^{0.68}_{-0.56}$ $R_{\oplus}$ and a period of 162.05 days.
    The highlighted regions are where the known candidate signal would transit the host star.
    Similar to Figure \ref{fig:percent_flagged}, we require a timestamp to be within at least 100 light curve segments to be included in the figure.
    Many of the transits appear near a data gap and hence influence the frequency they appear in a segment. In spite of the data gaps, we are still able to recover 6 of the 8 visible transits of KOI 1271.01, the second to last transit is completely in a data gap and is not visible at all. We recovered a period of 163.43 days, which is close to the known candidate period of 162.05 days. We also recover a transit of the new candidate, corresponding to the peak at day 679 BKJD.}
    \label{fig:KIC8631160_flagged}
\end{figure}

In early testing of this methodology, we wanted to test the effectiveness of our pipeline on systems outside our training, validation, and testing set. 
We randomly selected a handful of planets with a period greater than 100 days, and a transit depth larger than 350 ppm. 
This downselection ensures that we are outside the training regime since its cutoff was a maximum depth of 350 ppm while also still in the long orbital-period regime. 
One of the systems that were randomly selected was KOI-1271, which we focus on for the rest of this section. 
The host star has a radius of $1.48^{0.09}_{-0.07}$ $R_{\odot}$. 
This system contains one known candidate planet, KOI 1271.01, with a radius of $11.23^{0.68}_{-0.56}$ $R_{\oplus}$ and a period of 162.05 $\pm$ 0.00 days, the average SNR per transit of the planet is 131 \citep{Berger_2023, Holczer_2016}. 
KOI 1271.01 is known to have strong transit timing variations (TTVs), on the order of 100 minutes with a max TTV of 600 minutes; however, there is no other known transiting planet currently detected \cite{Ford2012, Santerne2016}. 
The percent flagged count plot, similar to Figure \ref{fig:percent_flagged}, of the KOI 1271 system is shown in Figure \ref{fig:KIC8631160_flagged}. 
The red highlighted regions are the locations where KOI 1271.01 transits the host star. 
The blue horizontal line is the noise level, which is 214.96.
The first three transits within \emph{Kepler} appear very close to data gaps and the second to last transit falls completely within a data gap. 
However, our pipeline is still able to recover the majority of the transits above our noise threshold. 
We recovered a period of 163.43 days, which is well within the bounds we choose to declare a successful recovery especially since this system has known strong TTVs. 

\begin{figure}
    \centering
    \includegraphics[width=8cm]{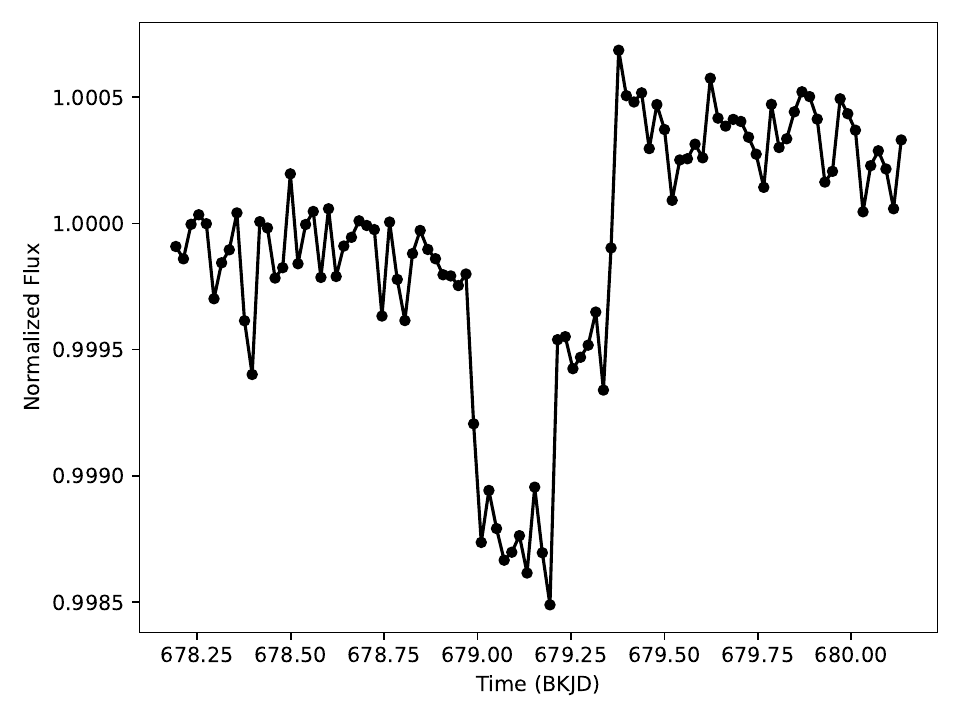}
    \caption{A zoom in the portion of the light curve causing the peak in Figure \ref{fig:KIC8631160_flagged} at day 679 BKJD. The total length shown is roughly 2 days, allowing 1.5 times the transit duration on both sides of the transit for a baseline flux. 
    The total time for the transit-like event is roughly 10 hours. We observe an offset near the center of the transit. The offset may be the reason why this transit went unnoticed.
    }
    \label{fig:1271.02_transit}
\end{figure}

Visible in Figure \ref{fig:KIC8631160_flagged} is a prominent peak in between the transits of the known candidate signal, at day 679 BKJD. 
The portion of KOI 1271's light curve that causes this peak is shown in Figure \ref{fig:1271.02_transit}.
We believe this signal is a single transit event of another planet within the system that is responsible for the strong TTVs of KOI 1271.01. 
We observe an offset in the middle of the transit and suggest this as a possible cause as to why it was unreported before this report.
We dedicate the rest of this section to analyzing this signal. 
For clarity between the known candidate planet KOI 1271.01, and the new signal found, we move forward calling the new candidate planet KOI 1271.02.

\subsection{Transit Fitting Potential Candidate} \label{subsec:MCMC_Transit}
The transit event from KOI 1271.02 happens on day 679.15 BKJD. 
This was found using the feature mapping technique, as described in \S\ref{subsec:feature_mapping}. 
Our pipeline only detected this signal once within \emph{Kepler}, making this a single transit detection. 
However, this also makes constraining the orbital and planetary parameters more difficult. 
We can make additional constraints on the period by observing where we do not see a transit, and utilizing that KOI 1271.01 has strong TTVs, see \S \ref{subsec:TTVs}. 
We assert that this strong of a TTV signal can only occur when the perturber is in a resonance configuration with the perturbed planet \citep{Agol_2005, Holman_2005}.
The resonances of the perturber to the perturbed that we investigated are 2:1; 3:1; 4:1; 5:1; 6:1; 7:1; 8:1; 9:1; 3:2; 5:2; 7:2; 9:2; 4:3; 5:3; 7:3; and 7:5. 
By checking the locations within \emph{Kepler} of where another transit of KOI 1271.02 would appear, we conclude that the only viable resonances not ruled out with extant \emph{Kepler} data are 7:2, 5:1, 6:1, 7:1, 8:1, and 9:1. 
These resonances either do not produce any other transits that would appear in \emph{Kepler} or happen close enough to a data gap that if the perturber experienced its own TTV, the transit could be moved into the data gap. 
The other resonances would have the perturber appear multiple times within the \emph{Kepler} dataset, where they are not observed. 

Since we only detect a single transiting event for the new candidate, we can perform a transit fit over a small range of times around the transit. 
We take 1.5 times the transit duration on both sides of the transit as our baseline flux.
We perform an MCMC transit model fit over the roughly 2-day window. 

Within the single transit of KOI 1271.02 is a discontinuity, shown in Figure \ref{fig:1271.02_transit}. A similar discontinuity appears in other parts of the quarter; however, no other discontinuity contains a transit-like feature nor were they picked up by the neural network pipeline. 
If we fit the data without adjusting for the jump in data, we would preferentially prefer grazing transits and significantly larger planets. 
We therefore account for this discontinuity by applying a vertical shift to all points after the jump and detrending the light curve with the applied shift.

\begin{figure}
    \centering
    \includegraphics[width=8cm]{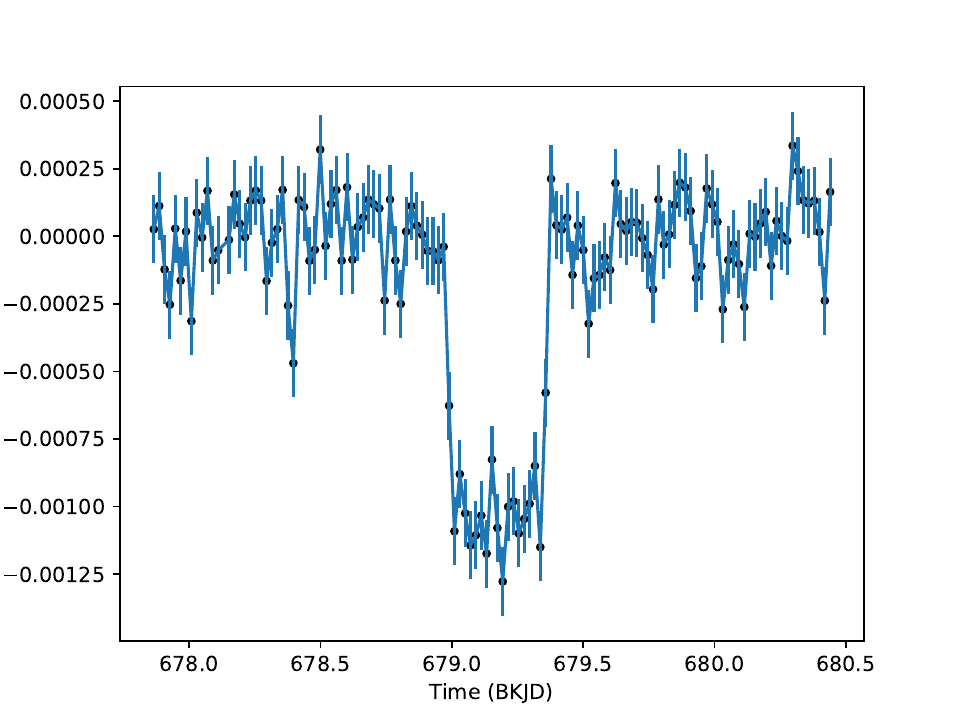}
    \caption{The detrended light curve segment of KOI 1271.02's single transit, with the applied best-fit offset from the MCMC. This segment is used for all of the transit-fitting MCMCs. The best-fit offset is $7.82 \times 10^{-4}$.}
    \label{fig:baseline}
\end{figure}

To find the best offset that should be applied, we perform an MCMC with only one parameter: the magnitude of the shift. 
Within the MCMC, after the shift is applied we detrend the entire light curve. 
We then grab a small window centered about the transit to inspect how well the detrending put the baseline flux about 0.
The MCMC minimizes a $\chi^2$ comparing the baseline flux outside the transit to 0.
The best offset after running the MCMC was $7.82 \times 10^{-4}$ in relative flux units. 
This offset was used for the remainder of the transit fitting, independent of the resonant period being fit. 
Figure \ref{fig:baseline} shows the final result of the light curve segment with the applied best-fit offset. 

We use \texttt{EXOPLANET} to model the 2-day transit window segment of the detrended light curve \cite{EXOPLANET}.
The parameters that we fit over are:\\
1) the period of the planet candidate, KOI 1271.02 \\
2) the time of central transit for KOI 1271.02 \\
3) the mass of KOI 1271 \\
4) the radius of KOI 1271 \\
5) the impact parameter of KOI 1271.02 \\
6) the eccentricity of KOI 1271.02 \\
7) the ratio of the planet to star radii \\
8 \& 9) the quadratic limb darkening parameters.\\
We use a uniform prior for the period centered on each resonance being tested and $\pm$ 2 days. 
We use a normal prior for the time of central transit centered about the observed peak at 679.15 BKJD and a standard deviation of 0.1 days.
For the stellar mass and radius, we use normal priors centered at 1.294 $M_{\odot}$ and 1.479 $R_{\odot}$, respectively, with a standard deviation of 0.01 \cite{Berger_2023}.
We hold uniform priors on the impact parameter and eccentricity to keep the system as a transiting system: impact parameter 0 to 1, and eccentricity from 0 to 0.4. The upper eccentricity bound was to ensure dynamical stability to first order and the two planetary orbits do not enter each other's hill radius. 
For the ratio of the planet to star radii, a uniform prior was held with lower and upper bounds of 0.001 and 0.2, respectively.
We hold normal priors about the quadratic limb darkening parameters \citep{Kipping_limb_dark, Agol_limb_dark}.
We run \texttt{EXOPLANET} with 54 chains for 8000 draws and 8000 tuning steps, for each resonance period. 
We talk about each resonance fit later in the section and also report all best-fit values with their respective uncertainty in Table \ref{tab:results}.

\subsection{TTVs of KOI 1271.01}\label{subsec:TTVs}
\begin{figure*}
    \centering
    \includegraphics[width = 17cm]{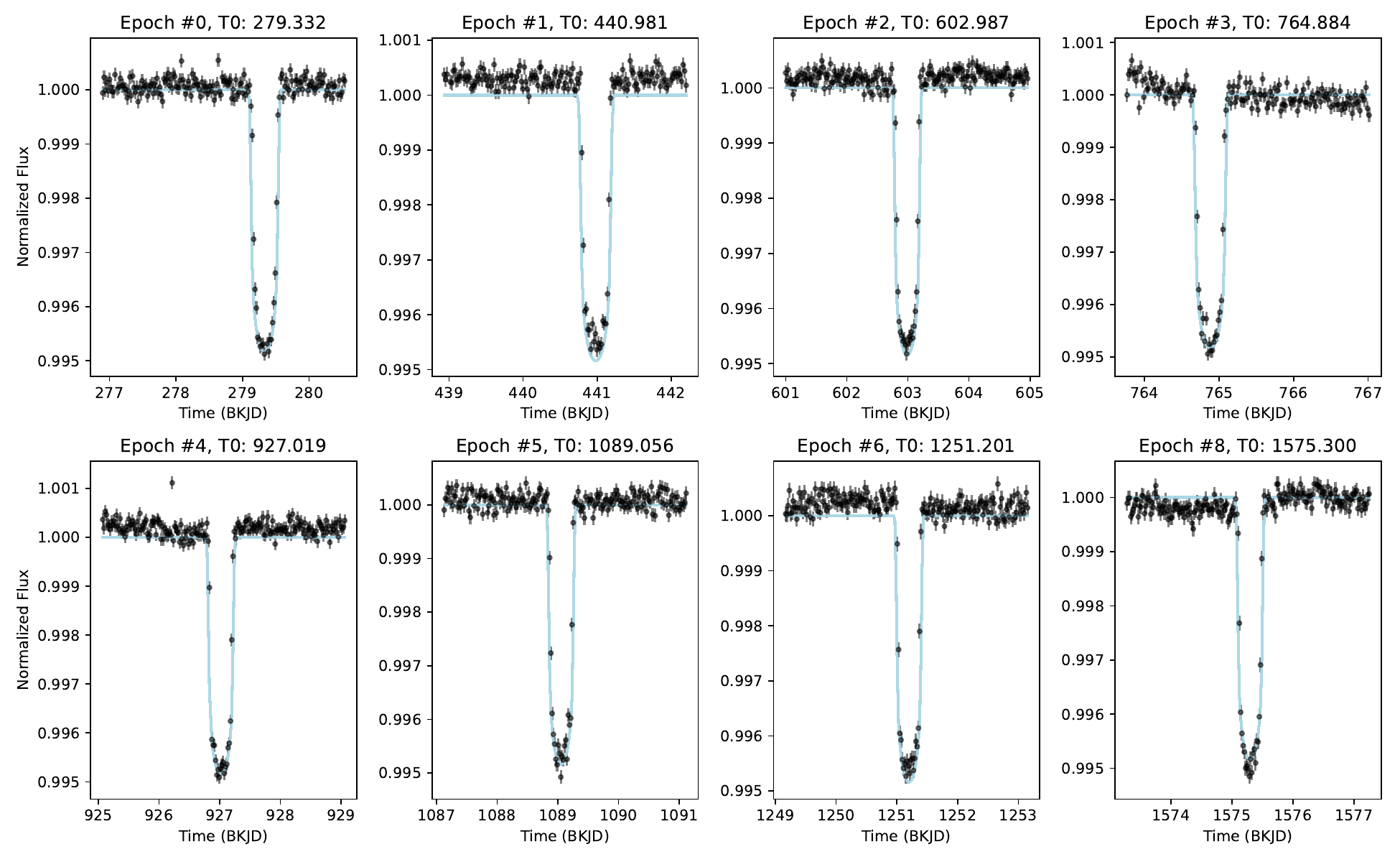}
    \caption{Best-fit transit models for all 8 of the visible transits within \emph{Kepler} of KOI 1271.01. Overlaid with the models is a scatter plot of the normalized flux values for the 4-day window with the associated error bars given by \emph{Kepler}. The models were found using \textsc{EXOPLANET}, and the only parameter that was fit for was the time of center transit. The title of each subplot is the epoch number along with the best-fit time of the transit center. We fit the models over a 4-day range around the predicted time of transit using the ephemeris of KOI 1271.01. Therefore, the location of the transit within the window gives a hint of the order of magnitude of the epoch's TTV.}
    \label{fig:1271_times}
\end{figure*}

KOI 1271.01 has TTVs on the order of 100 minutes in amplitude (\cite{Ford2012}, \cite{Santerne2016}). Here we perform our own analysis of the known TTVs of KOI 1271.01.
We first calculate our own period of KOI 1271.01. This was obtained by fitting a single transit on each individual transit of KOI 1271.01. The only parameter that was iterated over was the central time of the transit. This transit fitting was performed on all 8 visible transits within \emph{Kepler} of KOI 1271.01. We used \texttt{EXOPLANET} to vary over the central transit time while keeping all other parameters fixed to get a best-fit transit time for each individual transit \cite{EXOPLANET}. 
Figure \ref{fig:1271_times} shows all of the best-fit transit models with each respective observed transit. The center time in each panel of Figure \ref{fig:1271_times} is the predicted transit time from the linear ephemeris, therefore the horizontal location of the transit within each panel shows the strength of the TTV at that epoch.

\begin{figure}
    \centering
    \includegraphics[width=8cm]{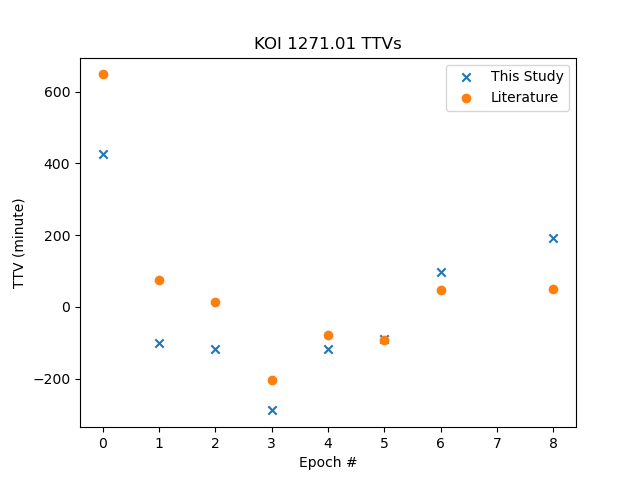}
    \caption{The TTVs for each transit for KOI 1271.01. The orange dots represent the difference between the calculated central transit time with P = 162.05 days, and T0 = 278.88 BKJD from \cite{Holczer_2016} and the observed times for KOI 1271.01.
    The blue 'x' points use the best-fit period and T0 from this study, P = 162.02 days and T0 = 279.04 BKJD, when performing the observed minus calculated central transit times. Epoch 7 falls into a data gap and is not observed. }
    \label{fig:1271_ttv}
\end{figure}

After a list of transit times was found, we performed a linear fit on the observed transit times. 
We recovered a best-fit period of 162.01 $\pm$ 0.02 days and a best-fit T0 of 279.04 $\pm$ 0.11 BKJD. 
Moving forward, we use these values for KOI 1271.01 instead of those reported in \cite{Batalha_2013}. 
We can now create an observed minus calculated (O-C) plot, and compare the values for the period and T0 from the literature and this study. 
Figure \ref{fig:1271_ttv} shows the O-C plot for the literature values \citep{Batalha_2013} and this study's best-fit values for period and T0. 
Our best-fit values decrease the TTV for the first transit to appear in \emph{Kepler}. 
It is important to note here that we do not observe a complete period of the TTV cycle for KOI 1271.01, due to the long orbital period of KOI 1271.01. 
The result from this is that our best-fit value for orbital period and T0 may not be accurate and the O-C plot may change with future observations of KOI 1271. 
This may also affect our dynamical fits to the system.  
Additional observations of the transits of KOI 1271.01 are needed to further constrain its period.

\subsection{Recreating TTV Signal of KOI 1271.01}\label{subsec:MCMC_TTV}
A single transit detection would normally only allow for very loose constraints on its orbital configurations, due to the degeneracy of the orbital parameters. 
However, in this case, there is already a known planet candidate within the system. 
Moreover, KOI 1271.01 experiences strong TTVs which provides us an opportunity to further constrain the orbital parameters of KOI 1271.02 by trying to reproduce the observed TTV signal, Figure \ref{fig:1271_ttv}.

We perform an MCMC to find the best orbital configuration and planetary parameters to reproduce the observed TTV signals of KOI 1271.01. We use \texttt{TTVFast} to produce the TTV signals of a two planetary system and parallelize \texttt{EMCEE} across 64 CPUs iterating over a total of 12 parameters:\\
1) mass of KOI 1271\\
2) mass of KOI 1271.01\\
3) eccentricity of KOI 1271.01 \\
4) inclination of KOI 1271.01 \\
5) argument of periastron for KOI 1271.01 \\
6) mean anomaly of KOI 1271.01 \\
7) mass of KOI 1271.02 \\
8) period of KOI 1271.02 \\
9) eccentricity of KOI 1271.02 \\
10) inclination of KOI 1271.02 \\
11) argument of periastron for KOI 1271.02 \\
12) mean anomaly of KOI 1271.02. \\

To produce an initial set of best-fit parameters, we perform an MCMC with very loose constraints.
We let the MCMC explore a 1 $\sigma$ deviation from the reported mass of 1.294 $M_{\oplus}$ \cite{Berger_2023}.
We let the MCMC explore a 10\% range about the mass of KOI 1271.01 as reported in \cite{Santerne2016}. Our priors for all angles are held at hard boundaries between 0 and 360 degrees. Our priors for the inclinations are such that they force the planet to be in a transiting configuration. 
The prior on the period of 1271.02 is that it must be within $\pm$ 5 days of the resonant period being tested. 
The prior for the mass of 1271.02 was held from 1 to 60 $M_{\oplus}$. 
Since the single transit candidate has a probable radius of 5.32 $\pm$ 0.33 $R_{\oplus}$, we found these bounds to provide plausible, albeit a little heavy, planet masses.
The large range for the mass prior allowed for the MCMC to explore lower eccentricities, as the mass and eccentricity are correlated in TTV fits.
The TTV signal is easier reproduced with high eccentricities; however, this results in unstable systems.
We therefore imposed strict constraints on the eccentricities, such that the outermost possible position of the inner planet was not within the Hill radius of the innermost possible position of the outer planet. 
The Hill radius is defined as:
\begin{linenomath}
\begin{align}
    R_{\rm Hill} &= a (1 - e) \sqrt[3]{\frac{m_p}{3\left(M_* + m_p \right)}},
\end{align}
\end{linenomath}
where $a$ is the semi-major axis in AU, $e$ is the eccentricity, $m_p$ is the planet mass in Solar masses, and $M_*$ is the stellar mass in Solar masses. 
For our priors, we choose the largest eccentricities such that:
\begin{linenomath}
\begin{align}
    \left[ a_b(1+e_b) + R_{\rm Hill}^b \right] &< \left[ a_c(1+e_c) + R_{\rm Hill}^c \right],
\end{align}
\end{linenomath}
where the sub/superscript $b$ denotes KOI 1271.01, and $c$ denotes KOI 1271.02.

For our log-likelihood function, we compute the square of the difference between the observed and calculated TTVs. 
If any of the epochs have more than a 60-minute, 2 \emph{Kepler} long cadences, difference between the calculated and observed TTV, we impose a penalty that adds 10 times the MAD times the sum of the difference of squares. 
The purpose of this step is to have the MCMC explore solutions that perform overall better on every single epoch, and not just fit a few transit times extremely well at the expense of being far off on others.

To obtain a valid starting point for our MCMCs, we do an initial coarse grid search to find a good starting place in the parameter space.
Initial testing when analyzing the coarse grid search showed that many possible combinations throughout the whole parameter space produce similar results.
Therefore, we cannot report for any resonance period a unique solution to its orbital configuration. 
For each resonance period, we simply report a possible configuration that is able to reproduce the observed large TTVs for which the MCMC converged.
For each solution in the parameter space, the values for the TTV at each epoch ended up at roughly the same position.
Therefore, showing an example solution is indicative of the best that resonance can do with the current data.

After this initial MCMC explored our parameter space, we ran an additional MCMC with tighter priors designed to explore the \texttt{best} solution for a given resonance rather than to broadly sample islands of solutions. This allows us to obtain realistic error bars for specific solutions after broadly quantifying the spread of possible solutions within a given resonance.

\subsection{Resonance 7:2}
We ran the described tests in \S\ref{subsec:MCMC_Transit} and \S\ref{subsec:MCMC_TTV} on the 7:2 resonance period (567.06 days).
We are able to accurately fit a transit model to the observed dip in the light curve holding the period fixed to within a few days of 567 days.
The ratio of the planet-to-star radius has a best-fit value of 0.033 $\pm$ 0.001 from the MCMC. 

\begin{figure}
    \centering
    \includegraphics[width=8cm]{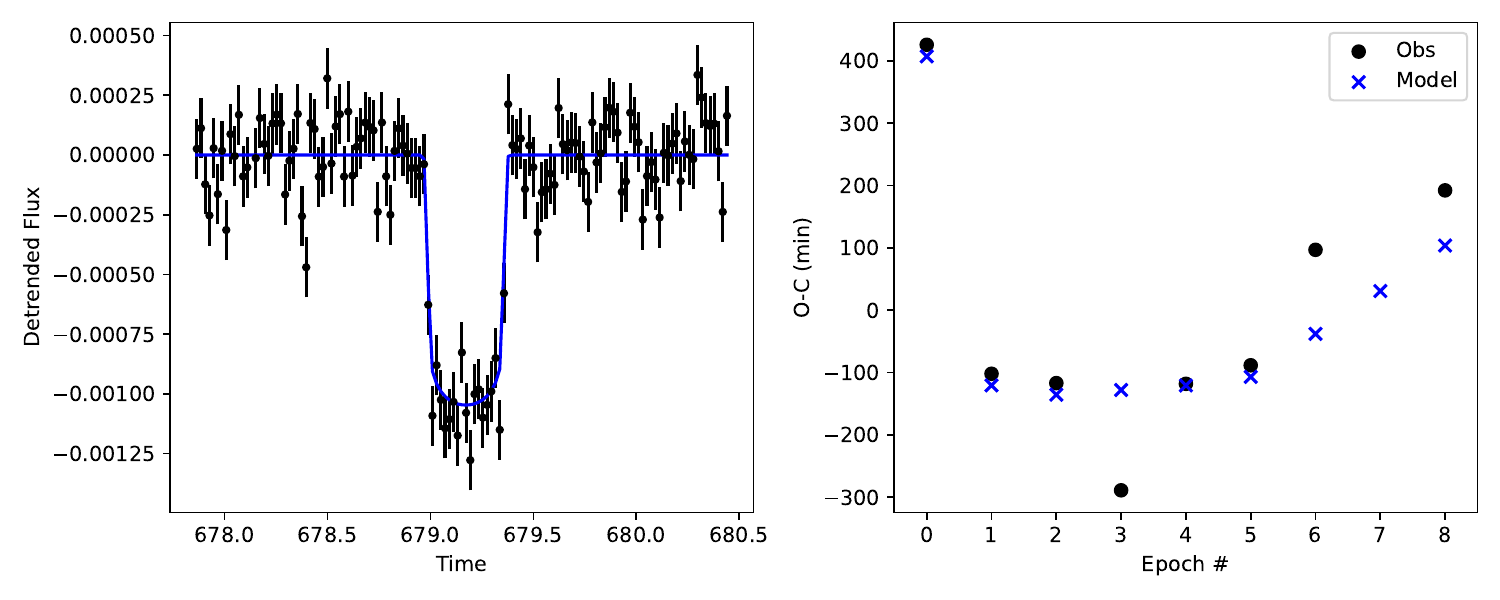}
    \caption{
    \emph{Left}) The best-fit 7:2 resonance transit model (solid blue) overlaid on an error bar plot of the detrended offset flux 2-day window (black). The transit model produces a radius of 5.32 $\pm$ 0.33 $R_{\oplus}$ for KOI 1271.02.
    \emph{Right}) The observed TTV signal of KOI 1271.01 (black dot) with a final configuration from an MCMC run with the 7:2 resonance period (blue x). The 7:2 resonance is able to fit epochs 0 \& 1 but struggles to fit epochs 2 \& 3.}
    \label{fig:res_72}
\end{figure}

With a 1.479 $\pm$ 0.01 $R_{\odot}$ star from the fit, this equates to a planet of radius 5.32 $\pm$ 0.20 $R_{\oplus}$. 
The TTV fit produced a mass of $28.943^{0.228}_{-0.470}$ $R_{\oplus}$ for KOI 1271.02.
This was also the best-fit mass value for the 5:1, 6:1, and 7:1 resonances.
The 7:2 resonance is the only resonance that is able to reproduce the observed TTV at epoch 0, as well as reproduce epoch 1.
This solution, however, struggles to reproduce the TTV signal in epochs 3, 6, and 7. 
We suggest this struggle is due to a lack of precision in the period of KOI 1271.01, and the lack of observing a full period of the TTV signal. 
We also note that the 7:2 resonance provided a solution with the lowest eccentricity for KOI 1271.01, $0.009^{0.002}_{-0.005}$.
The 7:2 resonance is currently our favored solution. Refinement of this solution will require additional transit observations of KOI 1271.01 and KOI 1271.02.
The final results for the 7:2 resonance are shown in Figure \ref{fig:res_72}.
The best-fit parameters and their uncertainty for the transit model and an example orbital configuration that produces the shown TTVs in Figure \ref{fig:res_72} are listed in Table \ref{tab:results}.

\subsection{Resonance 5:1}
\begin{figure}
    \centering
    \includegraphics[width=8.5cm]{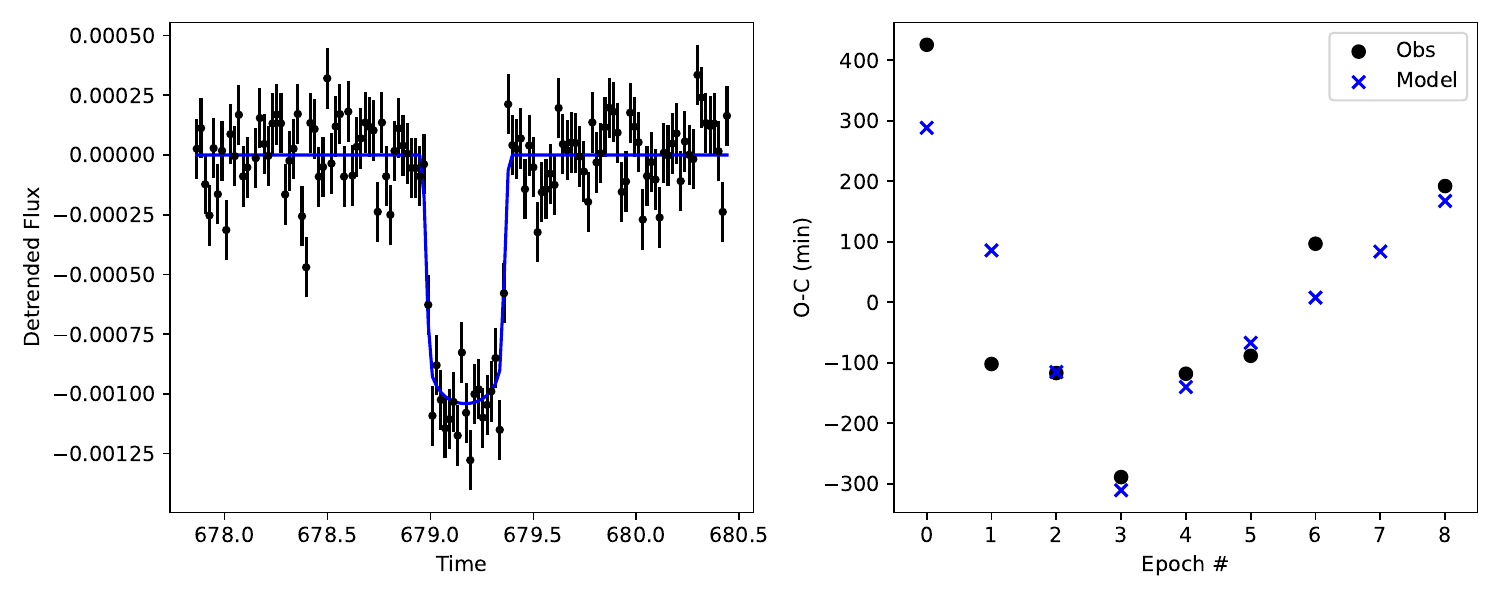}
    \caption{\emph{Left}) The best-fit 5:1 resonance transit model (solid blue) overlaid on an error bar plot of the detrended offset flux 2-day window (black) for KOI 1271.02. The best-fit radius is 5.32 $\pm$ 0.20 $R_{\oplus}$.
    \emph{Right}) The observed TTV signal of KOI 1271.01 (black dot) with a final configuration from an MCMC run with the 5:1 resonance period (blue x). The MCMC produces a mass of $42.094^{0.246}_{-1.075}$ $M_{\oplus}$ for KOI 1271.02. The 5:1 resonance is able to fit epochs 2, 3, 4, 5, \& 8 but fails to fit the chopping signal.}
    \label{fig:res_5}
\end{figure}

We performed two MCMCs on the 5:1 resonant period, roughly 810 days, for the best-fit transit model and a possible orbital configuration for the TTV signal. 
As seen in Figure \ref{fig:res_5}, we are able to fit a transit model with a 5:1 resonance period to the observed dip in the light curve.
Similar to the 7:2 resonance, the best-fit radius of this planet is 5.32 $\pm$ 0.20 $R_{\oplus}$.

To an extent, we are also able to reproduce the two largest TTV signal epochs. 
We found within our testing that the 5:1 resonance period struggles to reproduce the chopping observed in the TTV signal while also maintaining the large deviations in the transit timings. 
Chopping occurs on the time from one conjunction to the next conjunction, when their gravitational attraction is at maximum, and introduces a deviation in the TTV away from the regular periodic signal \citep{Agol_Fab_2018, Deck_2015}.
We found that we could reproduce the TTV signal with a wide range of large masses for KOI 1271.02, but show an example solution with a slightly large mass of $42.094^{0.246}_{-1.075}$ $M_{\oplus}$. 
We still require large eccentricities for KOI 1271.01 and KOI 1271.02, $0.414^{0.004}_{-0.006}$ and $0.394^{0.004}_{-0.002}$ respectively, to reproduce the large TTV signals of KOI 1271.01 for the 5:1 and higher resonances.
If we altered our priors to enforce a lower mass of KOI 1271.02, the MCMC was driven to higher eccentricities, as the TTV signal requires the product of the mass and eccentricity to be large to produce a large TTV \cite{Agol_Fab_2018}.
Therefore, the shown solution in Figure \ref{fig:res_5} and laid out in Table \ref{tab:results} is an example solution with a moderate mass and eccentricity.

\subsection{Resonance 6:1}
\begin{figure}
    \centering
    \includegraphics[width=8cm]{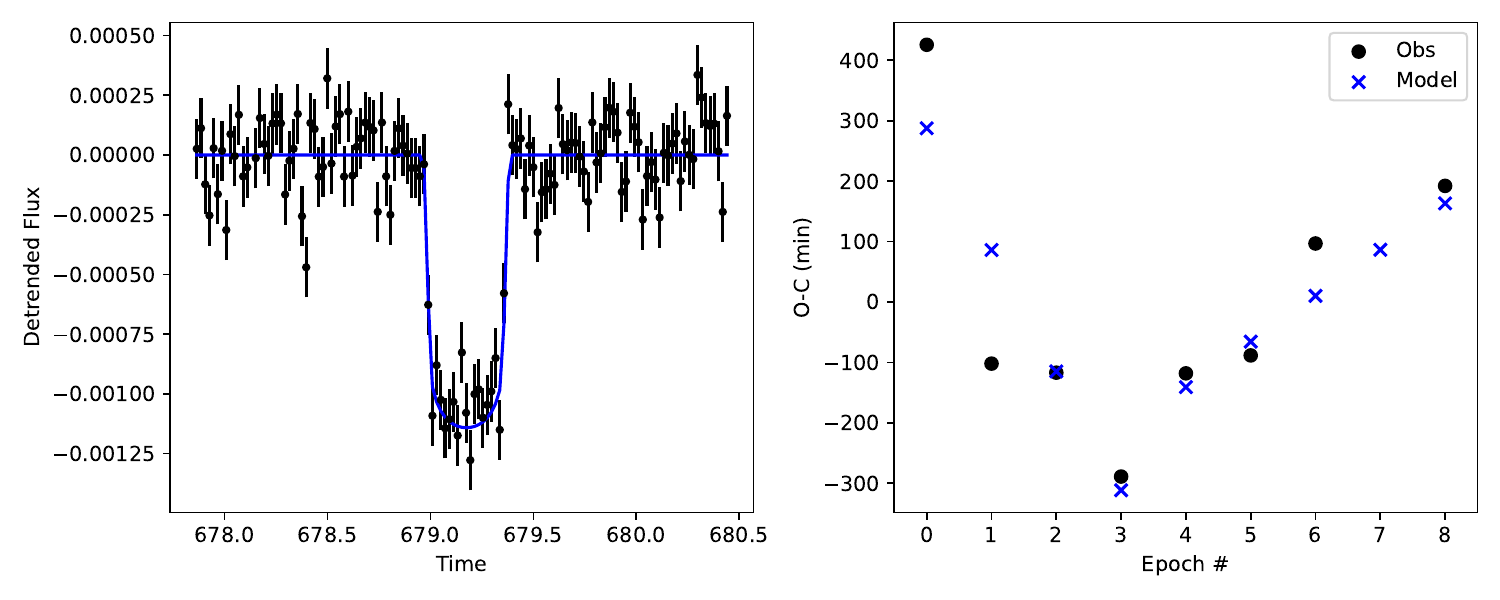}
    \caption{Holding the period fixed to within a few days of the 6:1 resonance period, 972 days, we present the best-fit transit model and an example solution for reproducing the observed TTV signal. 
    \emph{Left}) Best-fit transit model (blue) shown against an errorbar plot of the detrended, offset observed flux. The fitted radius is 5.32 $\pm$ 0.20 $R_{\oplus}$.
    \emph{Right}) Example of recreating the observed TTV signal (black dots) with an MCMC run (blue x). 
    The MCMC produces a mass of $31.807^{0.339}_{-0.177}$ $M_{\oplus}$ for KOI 1271.02.
    }
    \label{fig:res_6}
\end{figure}

We performed two MCMCs on the 6:1 resonant period or roughly 972 days for the best-fit transit model and a possible orbital configuration for the TTV signal. 
As seen in Figure \ref{fig:res_6}, we are able to fit a transit model with a 6:1 resonance period to the observed dip in the light curve. 
The best-fit radius is 5.32 $\pm$ 0.20 $R_{\oplus}$.

To an extent, we are also able to reproduce the two largest TTV signal epochs. We found within our testing that the 6:1 resonance period struggles to reproduce the chopping observed in the TTV signal while also maintaining the large deviations in the transit timings. 
We found that we could reproduce the TTV signal with a wide range of large masses for KOI 1271.02, but show an example solution with a reasonable mass of $31.807^{0.339}_{-0.177}$ $M_{\oplus}$.
The shown solution in Figure \ref{fig:res_6} and laid out in Table \ref{tab:results} is an example solution with a moderate mass and large eccentricities for KOI 1271.01 and KOI 1271.02, $0.425^{0.009}_{-0.008}$ and $0.497^{0.003}_{-0.004}$.

\subsection{Resonance 7:1}
\begin{figure}
    \centering
    \includegraphics[width=8cm]{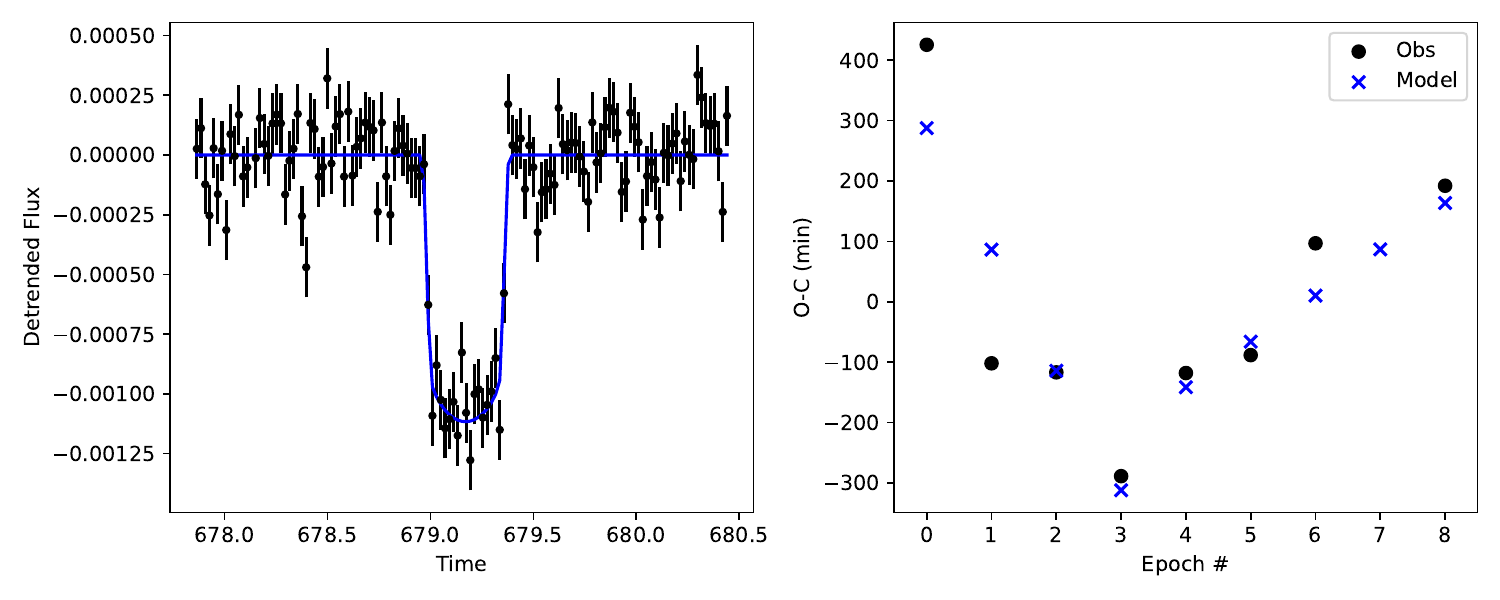}
    \caption{\emph{Left}) The best-fit 7:1 resonance transit model (solid blue) overlaid on an error bar plot of the detrended offset flux 2-day window (black). The radius is 5.32 $\pm$ 0.20 $R_{\oplus}$.
    \emph{Right}) The observed TTV signal of KOI 1271.01 (black dot) with a final configuration from an MCMC run with the 7:1 resonance period (blue x). KOI 1271.02 is found to have a mass of $36.957^{0.021}_{-0.045}$ $M_{\oplus}$.
    Similar to the 5:1 and 6:1, the 7:1 resonance can fit epochs 2, 3, 4, 5, and 8 but fails to fit the chopping signal.}
    \label{fig:res_7}
\end{figure}

Performing the tests on the 7:1 resonance resulted in the plots shown in Figure \ref{fig:res_7}.
The best-fit radius is the same as the 7:2, 5:1, and 6:1 resonances of 5.32 $\pm$ 0.20 $R_{\oplus}$.
Similar to the resonances 5:1 and 6:1, the 7:1 resonance is able to fit epoch 3 and get close to the extreme TTV of epoch 0. However, it fails to fit the chopping signal of epoch 1. 
The 7:1 resonance fits the TTVs with a reasonable mass for KOI 1271.02 of $36.957^{0.021}_{-0.045}$ $M_{\oplus}$ but requires high eccentricities for KOI 1271.01 and KOI 1271.02, $0.426^{0.002}_{-0.017}$ and $0.526^{0.003}_{-0.001}$ respectively.

\subsection{Resonance 8:1}
The 8:1 resonance was the first resonance that differentiated in the best-fit value for the ratio of planet to star radii. 
The best-fit radius of KOI 1271.02 for the 8:1 resonance is 5.49 $\pm$ 0.20 $R_{\oplus}$.
The radius is slightly larger than the shorter resonance periods' best-fit radii.
However, they are still within $\pm$1$\sigma$ of one another.

\begin{figure}
    \centering
    \includegraphics[width=8cm]{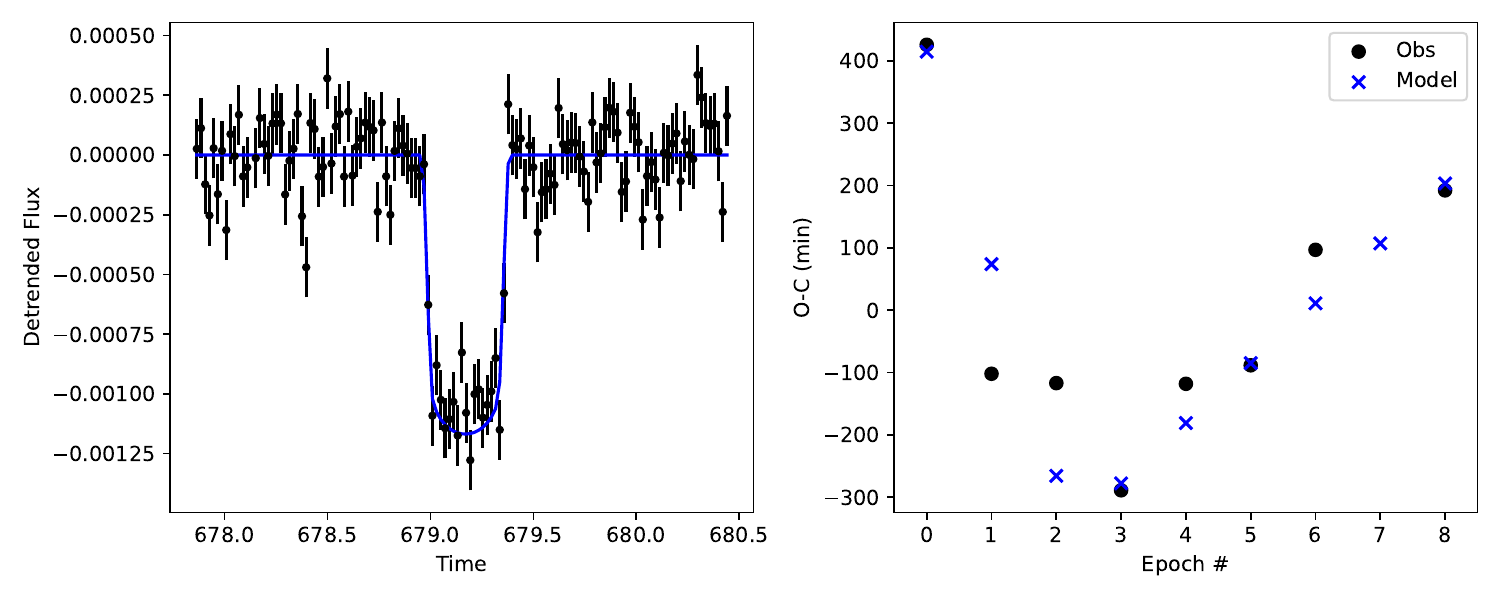}
    \caption{\emph{Left}) The best-fit 8:1 resonance transit model (solid blue) overlaid on an error bar plot of the detrended offset flux 2-day window (black). The predicted radius for KOI 1271.02 is 5.49 $\pm$ 0.20 $R_{\oplus}$.
    \emph{Right}) The observed TTV signal of KOI 1271.01 (black dot) with a final configuration from an MCMC run with the 8:1 resonance period (blue x). The TTV signal requires KOI 1271.02 to be $61.634^{0.240}_{-0.804}$ $M_{\oplus}$ for the 8:1 resonance.
    This resonance appears to be close to fitting some sort of chopping signal within the TTV, but cannot match the magnitude of the signal. The 8:1 resonance also fits epoch 0 and epoch 4 extremely well.}
    \label{fig:res_8}
\end{figure}

The 8:1 resonance appears promising as it appears to be producing a chopping signal that takes the form of the observed KOI 1271.01 chopping. 
However, it fails to meet the magnitude of the chopping signal in epochs 1 and 2. 
Perhaps one of the islands in parameter space would be able to match the signal exactly, but we were unsuccessful in finding such a solution. 
The 8:1 resonance had a low eccentricity for KOI 1271.01, $0.086^{0.001}_{-0.001}$, but required a high eccentricity of KOI 1271.02 with an eccentricity of $0.658^{0.002}_{-0.000}$, for the solution we propose. 
The solutions for the 8:1 resonance are shown in Figure \ref{fig:res_8}. 
The 8:1 resonance is the only resonance that fits the two extremes in the TTV signal, with epochs 0 and 4 being fit extremely well.
The 8:1 resonance produces a relatively large mass for KOI 1271.02 of $61.634^{0.240}_{-0.804}$ $M_{\oplus}$.

\subsection{Resonance 9:1}
\begin{figure}
    \centering
    \includegraphics[width=8cm]{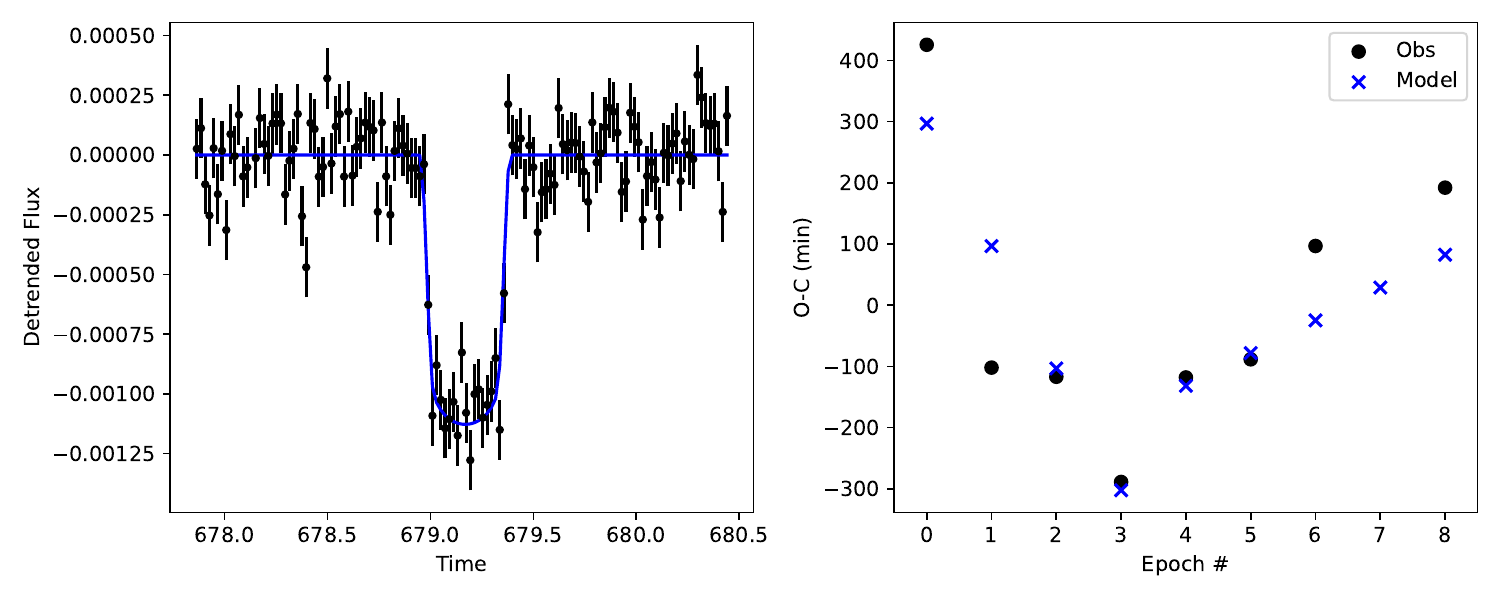}
    \caption{\emph{Left}) The best-fit 9:1 resonance transit model (solid blue) overlaid on an error bar plot of the detrended offset flux 2-day window (black). The best-fit radius of KOI 1271.02 is 5.49 $\pm$ 0.20 $R_{\oplus}$.
    \emph{Right}) The observed TTV signal of KOI 1271.01 (black dot) with a final configuration from an MCMC run with the 9:1 resonance period (blue x). 
    The best-fit mass for KOI 1271.02 is $32.708^{0.344}_{-1.035}$ $M_{\oplus}$.
    Similar to the 5:1, 6:1, and 7:1, the 9:1 resonance can fit epochs 2, 3, 4, and 5, but fails to fit the chopping signal at epochs 1 and 6.}
    \label{fig:res_9}
\end{figure}

Performing the tests on the 9:1 resonance resulted in the plots shown in Figure \ref{fig:res_9}.
The 9:1 produces the same solution for the ratio of the planet-to-star radius of 0.034 $\pm$ 0.001, making the radius of the planet 5.49 $\pm$ 0.20 $R_{\oplus}$.
Similar to the resonances 5:1, 6:1, and 7:1, the 9:1 resonance is able to fit epoch 3 and get close to the extreme TTV of epoch 0. However, it fails to fit the chopping signal of epoch 1. 
The 9:1 resonance fits the TTVs with a reasonable mass for KOI 1271.02 of $32.708^{0.344}_{-1.035}$ $M_{\oplus}$. 
The 9:1 resonance is able to produce the large TTVs of KOI 1271.01 with only a high eccentricity for KOI 1271.02, $0.671^{0.005}_{-0.001}$,  and a relatively low eccentricity for KOI 1271.01, $0.130^{0.000}_{-0.017}$, (compared to the other resonances).

\begin{table*} 
    \centering
    \begin{tabular}{|c||c|c|c|c|c|c|}
    \hline
    Resonance       &   7:2                 &       5:1             &      6:1              &    7:1                &         8:1           &         9:1 \\
    \hline
    Period & 567 $\pm$ 6 & 810 $\pm$ 6 & 972 $\pm$ 6 & 1134 $\pm$ 6 & 1296 $\pm$ 6 & 1458 $\pm$ 6 \\
    $T_0$ & 679.172 $\pm$ 0.002 & 679.171 $\pm$ 0.002 & 679.171 $\pm$ 0.002 & 679.171 $\pm$ 0.002 & 679.171 $\pm$ 0.002 & 679.171 $\pm$ 0.002 \\
    $M_*$ & 1.294 $\pm$ 0.01 & 1.294 $\pm$ 0.01 & 1.294 $\pm$ 0.01 & 1.294 $\pm$ 0.01 & 1.294 $\pm$ 0.01 & 1.294 $\pm$ 0.01 \\
    $R_*$ & 1.479 $\pm$ 0.01 & 1.479 $\pm$ 0.01 & 1.479 $\pm$ 0.01 & 1.479 $\pm$ 0.01 & 1.479 $\pm$ 0.01 & 1.479 $\pm$ 0.01 \\
    b & 0.738 $\pm$ 0.06 & 0.787 $\pm$ 0.037 & 0.807 $\pm$ 0.029 & 0.823 $\pm$ 0.023 & 0.837 $\pm$ 0.019 & 0.847 $\pm$ 0.015 \\
    e & 0.381 $\pm$ 0.091 & 0.413 $\pm$ 0.069 & 0.429 $\pm$ 0.058 & 0.441 $\pm$ 0.049 & 0.448 $\pm$ 0.044 & 0.457 $\pm$ 0.037 \\
    $R_p / R_*$ & 0.033 $\pm$ 0.001 & 0.033 $\pm$ 0.001 & 0.033 $\pm$ 0.001 & 0.033 $\pm$ 0.001 & 0.034 $\pm$ 0.001 & 0.034 $\pm$ 0.001 \\
    u1 & 0.284 $\pm$ 0.215 & 0.302 $\pm$ 0.226 & 0.306 $\pm$ 0.23 & 0.311 $\pm$ 0.235 & 0.307 $\pm$ 0.232 & 0.31 $\pm$ 0.238 \\
    u2 & 0.028 $\pm$ 0.203 & 0.01 $\pm$ 0.202 & 0.002 $\pm$ 0.197 & -0.001 $\pm$ 0.2 & -0.0 $\pm$ 0.197 & -0.003 $\pm$ 0.196 \\
    \hline
    $M_*$  &  $ 1.306 _{ -0.001 }^{ 0.000 } $ & $ 1.240 _{ -0.003 }^{ 0.005 } $ & $ 1.239 _{ -0.002 }^{ 0.010 } $ & $ 1.236 _{ -0.001 }^{ 0.003 } $ & $ 1.244 _{ -0.007 }^{ 0.003 } $ & $ 1.336 _{ -0.000 }^{ 0.001 } $ \\
    $M_b$  &  $ 240.912 _{ -0.089 }^{ 0.356 } $ & $ 260.609 _{ -0.275 }^{ 0.216 } $ & $ 251.094 _{ -0.195 }^{ 0.580 } $ & $ 244.054 _{ -0.542 }^{ 0.645 } $ & $ 273.110 _{ -0.523 }^{ 0.066 } $ & $ 288.800 _{ -0.215 }^{ 0.665 } $ \\
    $e_b$  &  $ 0.009 _{ -0.005 }^{ 0.002 } $ & $ 0.414 _{ -0.006 }^{ 0.004 } $ & $ 0.425 _{ -0.008 }^{ 0.009 } $ & $ 0.426 _{ -0.017 }^{ 0.002 } $ & $ 0.086 _{ -0.001 }^{ 0.001 } $ & $ 0.130 _{ -0.017 }^{ 0.000 } $ \\
    $i_b$  &  $ 89.433 _{ -0.082 }^{ 0.041 } $ & $ 89.643 _{ -0.075 }^{ 0.035 } $ & $ 89.586 _{ -0.049 }^{ 0.051 } $ & $ 89.991 _{ -0.039 }^{ 0.062 } $ & $ 90.101 _{ -0.041 }^{ 0.065 } $ & $ 90.154 _{ -0.053 }^{ 0.050 } $ \\
    $\rm{arg_b}$  &  $ 48.332 _{ -0.093 }^{ 0.326 } $ & $ 343.541 _{ -0.244 }^{ 0.301 } $ & $ 354.606 _{ -0.398 }^{ 0.618 } $ & $ 26.034 _{ -0.234 }^{ 0.474 } $ & $ 68.232 _{ -0.568 }^{ 0.478 } $ & $ 57.085 _{ -0.069 }^{ 1.901 } $ \\
    $\rm{anom_b}$  &  $ 12.388 _{ -0.325 }^{ 0.055 } $ & $ 268.879 _{ -0.443 }^{ 0.574 } $ & $ 251.088 _{ -0.854 }^{ 0.429 } $ & $ 208.510 _{ -0.374 }^{ 0.813 } $ & $ 36.819 _{ -0.379 }^{ 0.411 } $ & $ 1.806 _{ -1.286 }^{ 0.911 } $ \\
    $M_c$  &  $ 28.943 _{ -0.470 }^{ 0.228 } $ & $ 42.094 _{ -1.075 }^{ 0.246 } $ & $ 31.807 _{ -0.177 }^{ 0.339 } $ & $ 36.957 _{ -0.045 }^{ 0.021 } $ & $ 61.634 _{ -0.804 }^{ 0.240 } $ & $ 32.708 _{ -1.035 }^{ 0.344 } $ \\
    $P_c$  &  $ 572.225 _{ -0.036 }^{ 0.007 } $ & $ 810.006 _{ -0.065 }^{ 0.061 } $ & $ 968.175 _{ -0.051 }^{ 0.055 } $ & $ 1131.006 _{ -0.067 }^{ 0.063 } $ & $ 1295.893 _{ -0.060 }^{ 0.030 } $ & $ 1458.885 _{ -0.032 }^{ 0.089 } $ \\
    $e_c$  &  $ 0.511 _{ -0.002 }^{ 0.003 } $ & $ 0.394 _{ -0.002 }^{ 0.004 } $ & $ 0.497 _{ -0.004 }^{ 0.003 } $ & $ 0.526 _{ -0.001 }^{ 0.003 } $ & $ 0.658 _{ -0.000 }^{ 0.002 } $ & $ 0.671 _{ -0.001 }^{ 0.005 } $ \\
    $i_c$  &  $ 89.850 _{ -0.083 }^{ 0.309 } $ & $ 89.966 _{ -0.081 }^{ 0.124 } $ & $ 90.023 _{ -0.099 }^{ 0.089 } $ & $ 90.106 _{ -0.042 }^{ 0.036 } $ & $ 90.131 _{ -0.055 }^{ 0.021 } $ & $ 89.998 _{ -0.051 }^{ 0.046 } $ \\
    $\rm{arg_c}$  &  $ 212.109 _{ -0.032 }^{ 0.224 } $ & $ 141.277 _{ -0.146 }^{ 0.268 } $ & $ 136.067 _{ -0.681 }^{ 0.406 } $ & $ 181.444 _{ -0.285 }^{ 0.355 } $ & $ 261.111 _{ -0.286 }^{ 0.276 } $ & $ 223.874 _{ -0.065 }^{ 1.690 } $ \\
    $\rm{anom_c}$  &  $ 221.313 _{ -0.043 }^{ 0.105 } $ & $ 101.904 _{ -0.055 }^{ 0.047 } $ & $ 144.670 _{ -0.169 }^{ 0.103 } $ & $ 166.323 _{ -0.194 }^{ 0.120 } $ & $ 209.171 _{ -0.114 }^{ 0.140 } $ & $ 222.789 _{ -0.102 }^{ 0.123 } $ \\
    \hline
    \end{tabular}
    \caption{\emph{Top}) Best-fit values for each transit model for each resonance period along with their respected 1$\sigma$ standard deviation. 
    The ratio between the radius of the planet-to-star was found to be virtually the same across each resonance.
    For the 8:1 and 9:1  resonances, the resulting planet radius is 5.49 $\pm$ 0.20 $R_{\oplus}$.
    For all other resonances, the planet was found to have a radius of 5.32 $\pm$ 0.20 $R_{\oplus}$. 
    \emph{Bot}) An example orbital configuration that produces the large, observed TTVs of KOI 1271.01. The parameters with subscript $b$ denote belonging to KOI 1271.01. The parameters with subscript $c$ denote belonging to KOI 1271.02.}
    \label{tab:results}
\end{table*}

\subsection{Candidate Discussion}
We have now completed our MCMC runs fitting for the planetary parameters of KOI 1271.02 for each resonance that we believe are plausible configurations.
Our favored solution with the currently available data is the 7:2 resonance, as this is the only one that fits the TTVs of epochs 0 and 1. 
Moreover, the 7:2 resonance produced the TTV signals of KOI 1271.01 with the lowest eccentricity for KOI 1271.01, $0.009^{0.002}_{-0.005}$. 
Therefore, our favored solution would mean that KOI 1271.02 has a period of 567 $\pm$ 6 days, a radius of 5.32 $\pm$ 0.20 $R_{\odot}$, and a mass of $28.943^{0.228}_{-0.470}$ $M_{\oplus}$. 

\begin{figure}
    \centering
    \includegraphics[width=8cm]{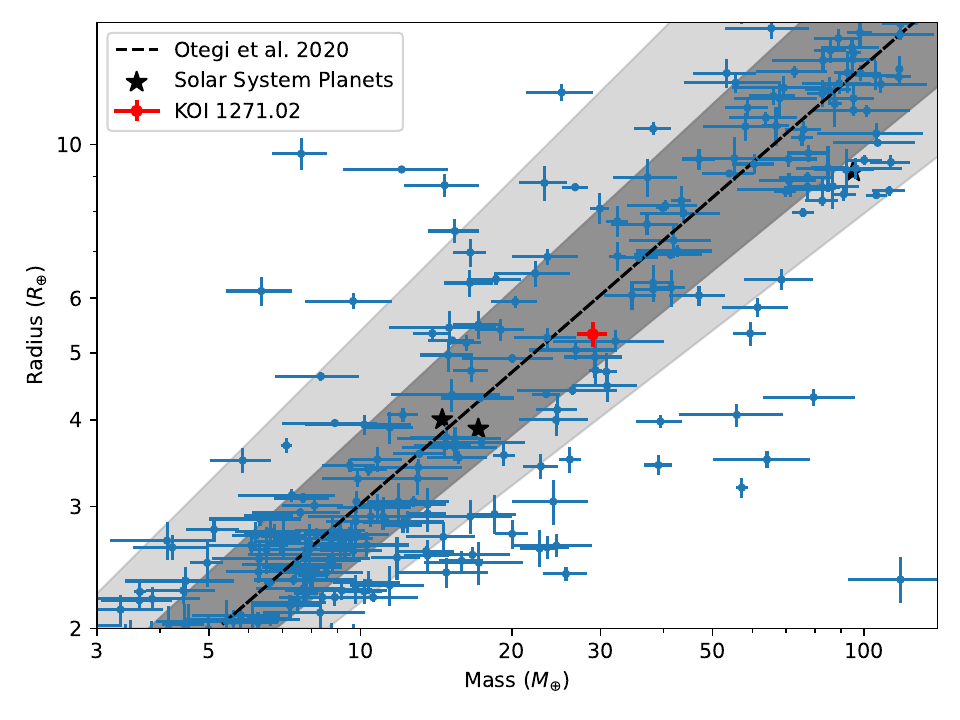}
    \caption{Mass-Radius diagram with known planets with relative uncertainties less than 25\% in mass, and 8\% in radius, along with Solar System examples, and our newly discovered candidate KOI 1271.02. The dashed line is the M-R power-law for volatile-rich exoplanets from \cite{Otegi_2020}. 
    The dark grey and grey-shaded regions correspond to the $\pm$1$\sigma$ and $\pm$2$\sigma$ of the fit, respectively.
    KOI 1271.02 is consistent within 1$\sigma$ of the volatile-rich power-law.
    This figure and selection criteria are adopted from \cite{Otegi_2020}. }
    \label{fig:mass_radius}
\end{figure}

Our TTV fitting produced a mass that is comparable to RV-fitted masses of known exoplanets.
Figure \ref{fig:mass_radius} shows a mass-radius relationship between observed exoplanets, along with Solar System planets, and the newly discovered KOI 1271.02.
The data for the known exoplanets is from the exoplanet catalog: Extrasolar Planets Encyclopaedia\footnote{\url{https://exoplanet.eu}}.
We adopt the same selection criteria as \cite{Otegi_2020} to produce the figure where the planetary mass must be below 120 $M_{\oplus}$, the relative uncertainty in the mass measurement must be less than 25\%, and the relative uncertainty in the radius must be less than 8\%.
With the same selection criteria, we adopt the power-law \cite{Otegi_2020} found for volatile-rich exoplanets: $M = (1.74\pm0.38) R^{1.58 \pm 0.10}$.
Our measurements of KOI 1271.02 found with transit and TTV fitting place KOI 1271.02 within 1$\sigma$ of their observed power-law.
KOI 1271.02 is consistent with a volatile-rich atmosphere composition.
Further observations of KOI 1271.01 or KOI 1271.02 will further constrain their orbital parameters and allow us to single out which resonance KOI 1271.02 may be in.

\section{TESS Follow Up}
With the currently available data in \emph{Kepler}, we have performed transit and TTV fits and produced a slew of solutions. 
One of the next steps in producing tighter constraints on the planetary parameters would be further observations of the KOI 1271 system.
As we do not see a complete TTV period of KOI 1271.01, we do not have tight constraints on its period. 
However, if we could observe another transit of KOI 1271.01 years after \emph{Kepler} we could further constrain the average period and help analyze its TTV signal.  

A portion of the \emph{Kepler} field was observed by \emph{TESS}. 
However, the period of KOI 1271.01 is 162 days, and a \emph{TESS} sector is only 28 days, so the probability of a transit of the known planet occurring in a \emph{TESS} sector is fairly low.
The period of KOI 1271.02 is even longer, and the signal of the transit is smaller, making the probability of a transit occurring within a sector as well as being able to detect it much lower.

\begin{figure}
    \centering
    \includegraphics[width=8cm]{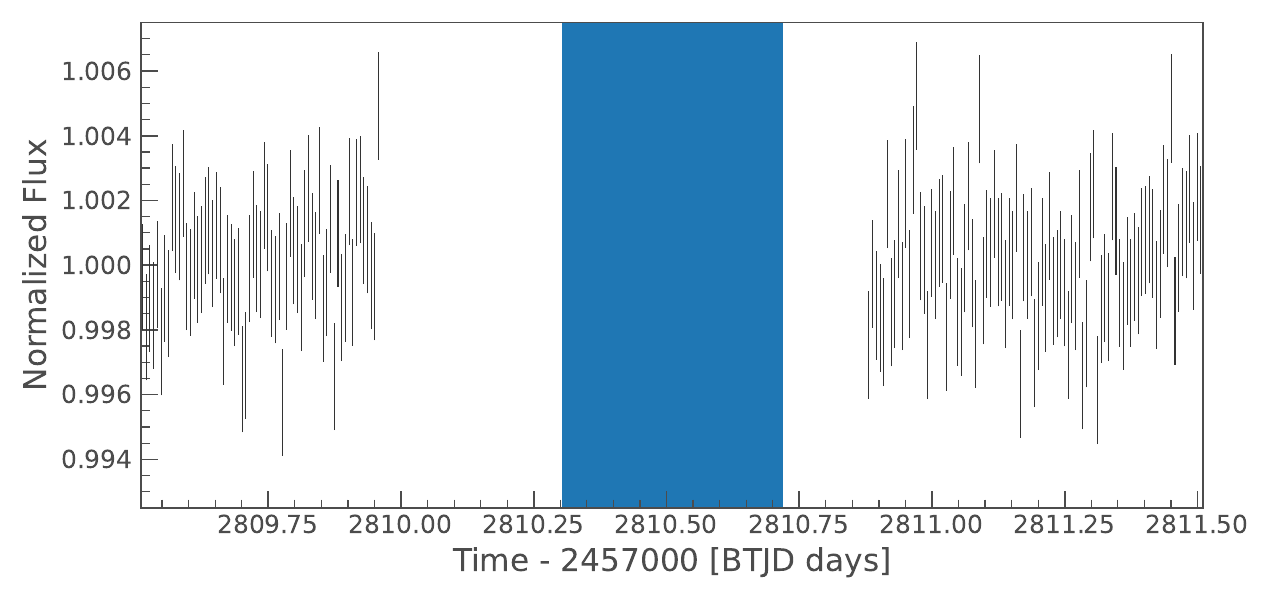}
    \caption{KOI 1271 follow up in \emph{TESS}. The highlighted blue region is the duration and location of where we predicted a transit of KOI 1271.01 to be. The only transit of KOI 1271.01 that occurs in a sector of \emph{TESS} takes place during a data gap.}
    \label{fig:tess_follow}
\end{figure}

KOI 1271 has appeared in \emph{TESS} in sectors 14, 15, 40, 41, 54, 55, and 74 and will appear again in sector 75.
We use the best-fit period for KOI 1271.01 and cast out predicted transit times within the time period of \emph{TESS}. 
A transit of KOI 1271.01 appears 5 days before sector 14 begins, and one appears in the middle of sector 55. The transit that occurs in sector 55 happens on day 2810.5 BTJD. 
However, when observing KOI 1271 in sector 55, \emph{TESS} had a data gap occur from roughly 2810 to 2811 BTJD. 
Therefore, the single time \emph{TESS} would have been observing KOI 1271 when a transit from KOI 1271.01 took place, there is a data gap. 
A zoom-in on sector 55 where we predict the transit to be is shown in Figure \ref{fig:tess_follow}. 
The highlighted region in blue is the duration of the transit.
Inspecting all sectors of \emph{TESS} that KOI 1271 appears in, we find no indication of any other transits within any of the sectors.

We used the resonance periods for KOI 1271.02 and projected future transits outwards within the time frame of \emph{TESS}, and no resonance period predicts a transit of KOI 1271.02 to be within a future \emph{TESS} sector that KOI 1271 is being observed in.

\section{Injection Recovery}\label{subsec:Inject_Rec}
The discovery of a potential new signal begs the question of what size planet our pipeline is sensitive enough to detect. To test the sensitivity of our pipeline, we perform an injection and recovery test of our end-to-end pipeline. We take KOI 1271's light curve, remove the known planets' signals, inject our own fake signals with a range of radii and periods, collect data, feed into our pipeline, and report if the pipeline recovered the signal and period. 

KOI 1271 has a $1.48^{0.09}_{-0.07}$ $R_{\odot}$ radius and has a known candidate planet, KOI 1271.01, with a radius of 11.52 $R_{\oplus}$ an orbital period of 162 days. This system is outside of the training, validation, and testing set that was used to train the ensemble. 
The first step we must take in producing an injection recovery test is removing all known signals from the light curve. Since KOI 1271.01 has known large transit timing variations (TTVs), we manually inspect each of the 8 transits to remove their signal. We also remove an unknown signal, the subject of \S\ref{sec:Candidate}, that produces a significant peak. The reason for removing the known peaks is so that the periodogram is picking up signals from the injection and not from the known planet. This will give a more accurate description of where the sensitivity of our pipeline for this star lies. 

After all the known peak-inducing signals are removed, we are able to start injecting fictitious signals into KIC 8631160's light curve. 
Using \texttt{BATMAN} we inject 2,923 independent planetary signals into KOI 1271 \cite{BATMAN}. 
For the first 1,000 injections, we ranged the planetary radius evenly in log space from 0.5 $R_{\oplus}$ to 16 $R_{\oplus}$ and the orbital period linearly from 50 days to 800 days. 
For the next 2,000 injections, we randomly sampled a planetary radius and orbital period between 0.5 $R_{\oplus}$ to 16 $R_{\oplus}$ and 50 days to 800 days, respectively.
For each injection, we created a data set and fed it into our pipeline, with no prior information on the location of the injections.
We do not adjust or edit the engineering attributes at all for the injection recovery test.
To declare a successful recovery, we require that at least 2 significant peaks were found and a period within 10 days of the injected period was returned. These criteria do not allow for a single transit detection to be classified as a successful recovery. We also do not check that all injected transits fall within the data, i.e. they are not injected within a data gap, to better simulate performance.

\begin{figure}
    \centering
    \includegraphics[width=8cm]{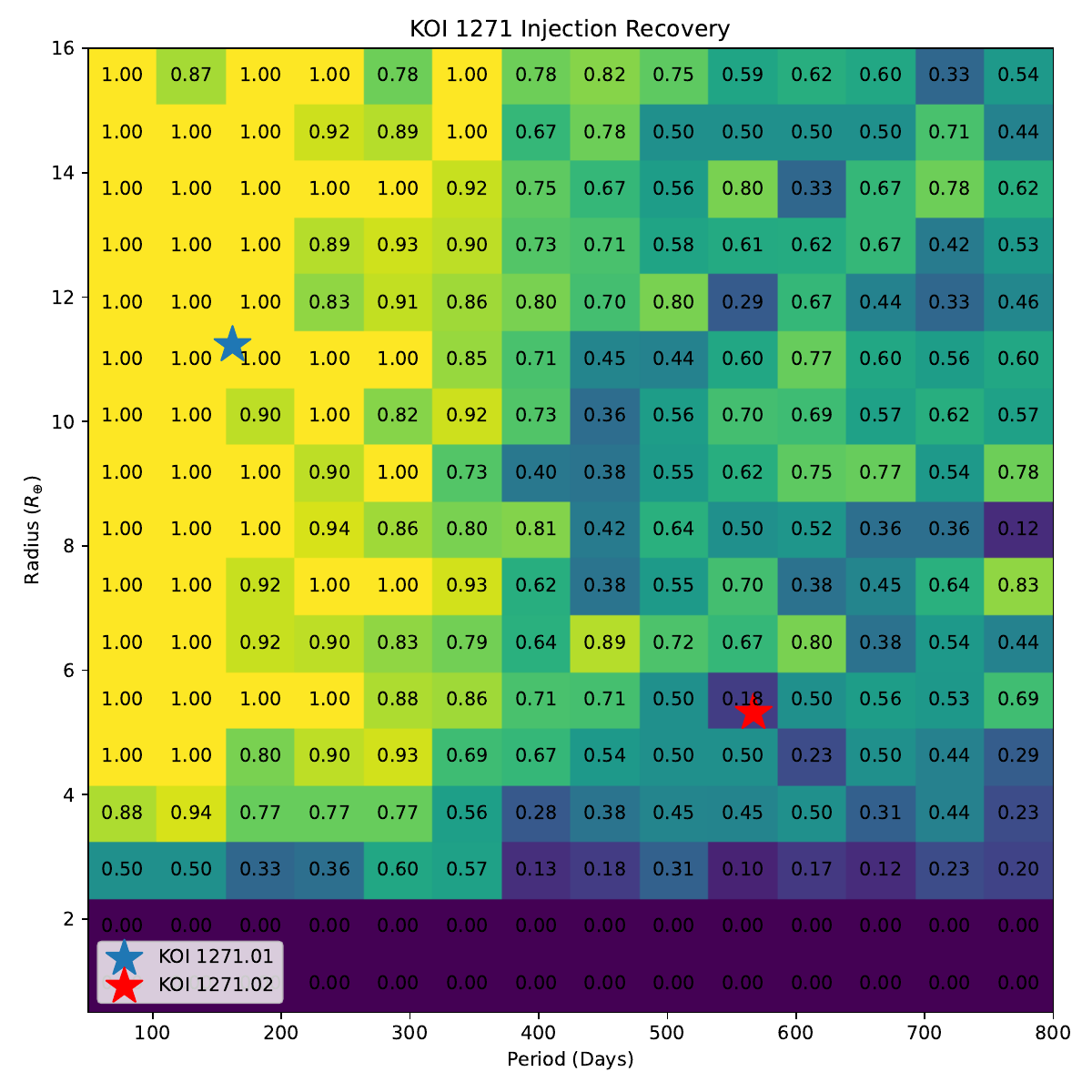}
    \caption{Injection recovery on star KOI 1271. We injected approximately 3,000 planetary signals with radii ranging from 0.5 to 16 $R_{\oplus}$, and an orbital period from 50 to 800 days. The numbers within the cells represent the fraction of successful recovers within that cell. We also place KOI 1271.01 and KOI 1271.02 within the plot for reference. For KOI 1271.02 we place the 7:2 resonance parameters. We define a successful recovery as having recovered at least 2 transits and the period within 10 days. Our sensitivity for planets around this 1.47 $R_{\odot}$ star is currently about 3.5 $R_{\oplus}$, and our pipeline preferentially recovers shorter orbital periods.}
    \label{fig:KIC8631160_inj_rec}
\end{figure}

Figure \ref{fig:KIC8631160_inj_rec} shows the injection recovery for KOI 1271. KOI 1271.01 has a radius of $11.23^{0.68}_{-0.56}$  $R_{\oplus}$ and an orbital period of 162 days. 
This is within the region where our pipeline is confident for recovery on this star. 
Due to the nature of our recovery requirements, our pipeline preferentially prefers shorter periods. 
KOI 1271 has approximately 1400 days of data, so it is not certain that an injection of 700 days or longer produces 2 available transits to be detected. 
The pipeline's sensitivity for a confident recovery is around 3.5 $R_{\oplus}$. 

\begin{figure}
    \centering
    \includegraphics[width=8cm]{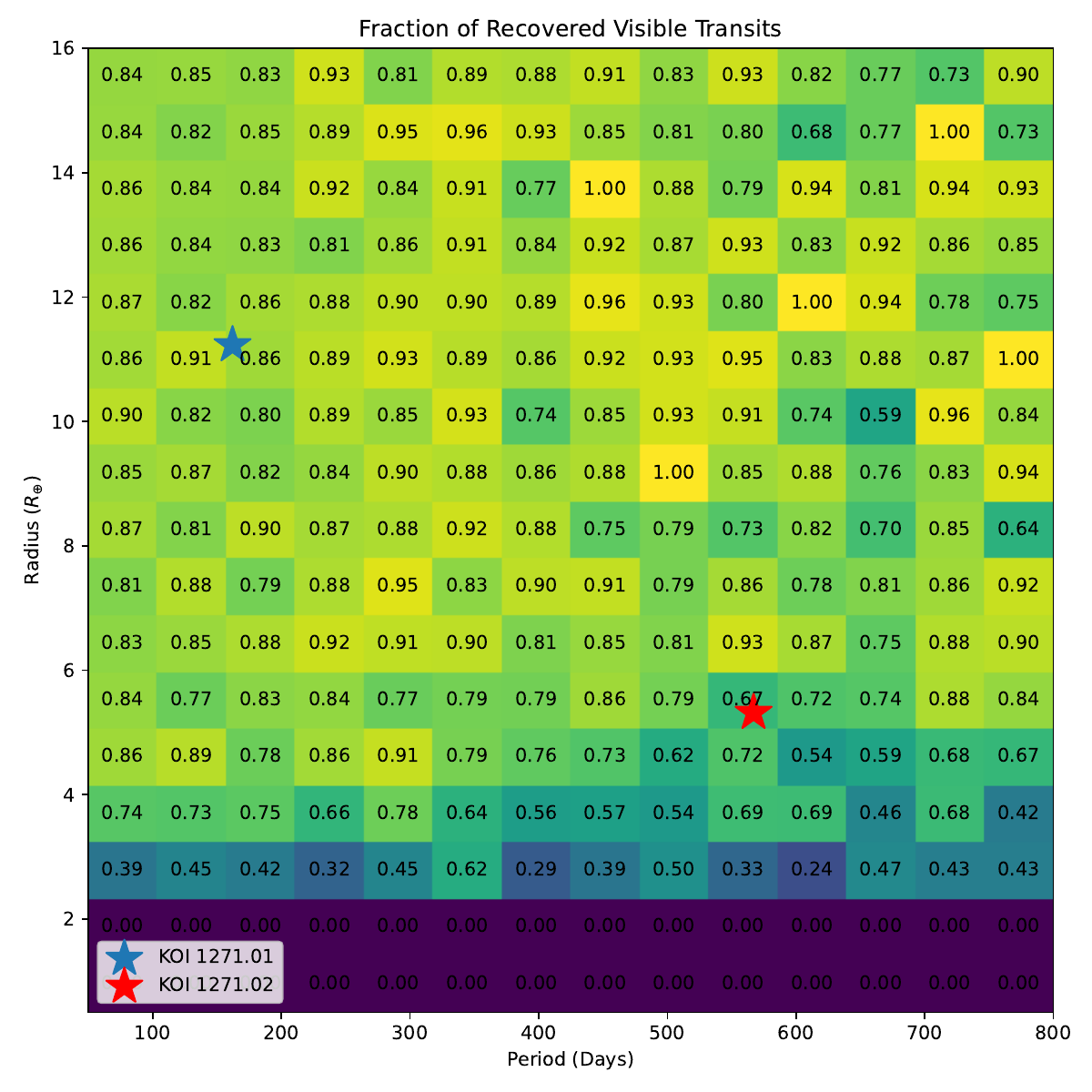}
    \caption{The fraction of recovered visible transits on star KOI 1271. These are the same injected signals as in Figure \ref{fig:KIC8631160_inj_rec}. Here, we report the ratio of the number of transits that were recovered to the number of transits that fall within the data, and not in a data gap. We also place KOI 1271.01 and KOI 1271.02 within the plot for reference. For KOI 1271.02 we place the 7:2 resonance parameters. We observe almost no dependence on orbital period for recovering transits. We do report that there is a strong gradient in the fraction of recovered transits from 4 to 2 $R_{\oplus}$. We therefore report that we are sensitive to single transits down to about 3 $R_{\oplus}$ for KOI 1271.}
    \label{fig:KOI1271_rec_trans}
\end{figure}

Due to the bias in how we define a successful injection recovery, we also present in Figure \ref{fig:KOI1271_rec_trans} the fraction of visible transits that are recovered. We state that a visible transit is a transit that does not fall into a data gap of the star. Figure \ref{fig:KOI1271_rec_trans} shows the total amount of transits recovered over the total amount of transits that fall within a section of data. Therefore, Figure \ref{fig:KOI1271_rec_trans} is a more accurate representation of the radius sensitivity of our pipeline, and we would expect to see almost no dependence on the orbital period of the injection. 

We have shown that we are able to recover extremely long-orbital period planets down to about 3 $R_{\oplus}$ around a large 1.47 $R_{\odot}$ star with the current version of this pipeline.
The pipeline's period dependence diminishes as we only require signal transit detections to count as a success, further highlighting the potential impact of our pipeline on producing more \emph{Kepler} candidate planets.
The injection recovery yields promising results and we hope continued development of this pipeline and utilization of the spacecraft engineering data will continue to uncover new planet signals.

\section{Conclusion} \label{sec:Conclusion}
\emph{Kepler} was a statistical mission with the goal of measuring the frequency of Earth-like planets around Sun-like stars, $\eta$-Earth.
However, due to the sensitivity of \emph{Kepler} and the inherent biases of detection techniques, studies were forced to extrapolate out towards the long orbital periods of Earth-like planets.
In order to further constrain the value of $\eta$-Earth, we must be able to confidently detect longer-orbital period planets. 
Single transit detection techniques will enable such detections and further constrain occurrence rates for long-orbital period planets.

In this work, we have developed a novel approach to classifying single transit events within the \emph{Kepler} data set.
We use an ensemble of 25 uniquely built CNNs trained on the photometry and the ancillary engineering data of \emph{Kepler}.
We demonstrated the accuracy of our ensemble down to the rocky planet regime and reported the discovery of a new candidate in the KOI 1271 system.
The main results may be summarized as follows:

\begin{enumerate}
    \item The ancillary engineering files contain relevant information for classifying a transit and machine learning techniques are able to learn how to incorporate them in predicting single transits.
    We incorporated the onboard spacecraft diagnostics of \emph{Kepler}, and found the 5 attributes that best contribute to classifying transits without too much of a sacrifice on computational time: the mean attitude error about the y- and x-axis; the standard deviation of the attitude error about the z- and y-axis; the state of reaction wheel \#1.

    \item We built a single transit detection pipeline that is able to locate the precise positions of transits and return the orbital period, with no prior information on the stellar parameters or the planet.
    We are able to recover the period within $\pm$ 0.1 days of the reported literature value of the period.

    \item  We performed an injection and recovery test on a 1.47 $R_{\odot}$ star. Our pipeline is confident in its recovery of an orbital period out towards 400 days and is still able to recover an 800-day orbital period. 
    We observe no period relation with recovering single transits with our pipeline and we are sensitive down to 3.5 $R_{\oplus}$ planets.

    \item Our pipeline has found a single transit candidate within the KOI 1271 system, denoted KOI 1271.02. This recently unreported candidate may be responsible for the observed strong TTVs of the known planet within the system, KOI 1271.01. 
    
    \item We perform a series of MCMCs to constrain the planetary parameters of KOI 1271.02 with the assumption that it must be in a mean-motion resonance with KOI 1271.01. Our favored solution is the 7:2 resonance, making KOI 1271.02 have a period of 567 $\pm$ 6 days, a radius of 5.32 $\pm$ 0.20 $R_{\oplus}$, and a mass of $28.943^{0.228}_{-0.470}$ $R_{\oplus}$. KOI 1271.02 is consistent with a volatile-rich atmosphere.
\end{enumerate}

Since we do not observe a complete period of the TTVs for KOI 1271.01, we have only loose constraints on the period and hence the TTV signals themselves. We would require further observation time to view transits of KOI 1271.01, and potentially KOI 1271.02, to further constrain their period, their TTV signals, and their orbital configurations.
KOI 1271.01 had a potential transit within a sector of TESS; however, the transit time fell into a data gap in the middle of the sector.
More observations of KOI 1271 are needed to further constrain its two planets KOI 1271.01 and KOI 1271.02.

MH would like to thank JD for donating countless hours of his time to help me become a more thoughtful and complete researcher. 
MH would like to thank the Dittmann research group, Dittmannia, for useful discussion. 
MH would like to give special recognition to Ben Capistrant, Billy Schap, and Sheila Sagear for their help with modeling planetary parameters.

\section*{Software}
The authors acknowledge the University of Florida Research Computing for providing computational resources and support that have contributed to the research results reported in this publication\url{http://www.rc.ufl.edu)}.
\texttt{astropy} \cite{ASTROPY}, \texttt{BATMAN} \cite{BATMAN}, \texttt{EMCEE} \cite{EMCEE}, \texttt{EXOPLANET} \cite{EXOPLANET}, \texttt{lightkurve} \cite{Lightkurve}, \texttt{SciPy} \cite{scipy}, \texttt{TTVFast} \cite{TTVFast}

\section*{Data availability}
All the {\it Kepler} data used in this paper can be found in MAST: \dataset[10.17909/T98304]{http://dx.doi.org/10.17909/T98304}.

\bibliography{main}{}
\bibliographystyle{aasjournal}

\end{document}